\documentclass[aps,
prd,
superscriptaddress,
floatfix,
showpacs,
showkeys,
preprint,
tightenlines,
nofootinbib,
amssymb,
amsmath]
{revtex4-1}
\usepackage{bm}
\usepackage[dvips]{color,graphicx}
\usepackage[figuresright]{rotating}
\usepackage[hidelinks]{hyperref}
\usepackage{supertabular}
\usepackage{setspace}
\newcommand{\ud}     {\mathrm{d}}
\newcommand{\gev}    {\mathrm{GeV}}
\newcommand{\Mev}    {\mathrm{MeV}}
\newcommand{\gevsq}  {\mathrm{GeV}^2}
\renewcommand{\Im}{\mathop{\mathrm{Im}}}
\renewcommand{\Re}{\mathop{\mathrm{Re}}}

\newcommand{\ceps}{\varepsilon}

\newcommand{\average}[1]{\left\langle{#1}\right\rangle}
\newcommand{\eq}[1]{Eq.(\ref{#1})}
\newcommand{\Eqs}[2]{Eqs.~(\ref{#1}) and (\ref{#2})}
\newcommand{\eqs}[1]{Eqs.(\ref{#1})}
\allowdisplaybreaks[4]

\begin{document}
%
%
\title{%
Nuclear parton distributions and the Drell-Yan process   
}

\author{S.~A.~Kulagin}
\email[]{kulagin@ms2.inr.ac.ru}
\affiliation{Institute for Nuclear Research of the Russian Academy of Sciences, Moscow 117312, Russia}
\author{R.~Petti}
\email[]{Roberto.Petti@cern.ch}
\affiliation{Department of Physics and Astronomy, University of South Carolina, Columbia SC 29208, USA}


\begin{abstract}
We study the nuclear parton distribution functions on the basis of  
our recently developed semi-microscopic model, 
which takes into account a number of nuclear effects including
nuclear shadowing, Fermi motion and nuclear binding, nuclear meson-exchange currents, and off-shell corrections
to bound nucleon distributions.
We discuss in detail the dependencies of nuclear effects on the type of parton distribution (nuclear sea vs valence),
as well as on the parton flavor (isospin).
We apply the resulting nuclear parton distributions to
calculate ratios of cross sections for proton-induced Drell-Yan production
off different nuclear targets.
We obtain a good agreement on the magnitude, target and projectile $x$, and the dimuon mass 
dependence of proton-nucleus Drell-Yan process data from the E772 and E866 experiments at Fermilab.
We also provide nuclear corrections for the Drell-Yan data from the E605 experiment.  
\end{abstract}

\preprint{INR-TH-2014-004}
\preprint{CETUP2014-002}

\maketitle

\section{Introduction}
\label{sec:intro}

Relying on the QCD factorization theorem \cite{Collins:1989gx},  
parton distributions (PDFs) determine the leading contributions 
to the cross sections of various hard processes involving leptons and
hadrons. In this context, PDFs are universal process-independent characteristics
of the target at high invariant momentum transfer $Q$, which are extracted from global fits
\cite{Alekhin:2012ig,Martin:2009iq,Ball:2012cx,Gao:2013xoa}
using data on lepton-nucleon deeply inelastic scattering (DIS), as well as the data on muon pair production in hadron
collisions (Drell-Yan reaction, or DY).

Electron and muon DIS experiments off nuclear targets 
demonstrated significant nuclear effects
with a rate that is more than one order of magnitude larger than the ratio of the nuclear 
binding energy to the nucleon mass \cite{Arneodo:1992wf,Norton:2003cb}.
These observations rule out the naive picture of the nucleus as a system of quasi-free nucleons and 
indicate that the nuclear environment plays an important role even at energies and momenta much higher 
than those involved in typical nuclear ground-state processes
\cite{Arneodo:1992wf,Norton:2003cb,Geesaman:1995yd,Bickerstaff:1989ch}.

A few phenomenological approaches to nuclear parton distributions (nPDFs) are available in the literature
\cite{Eskola:2009uj,Hirai:2007sx,deFlorian:2011fp}.
Typically, such analyses assume separate
nuclear corrections for each PDF, which are extracted from global fits to nuclear data including  
DIS, DY production, heavy-ion collisions, etc. Although these studies are useful in constraining nuclear effects
for different partons, they provide little information about the underlying physics mechanisms 
responsible for the nuclear corrections. Furthermore, they result in a large number of free parameters, as well 
as nuclear correction factors incorporating explicit parametrizations of the nuclear dependence. 
We also note that the current phenomenology of nuclear effects in neutrino DIS leads to somewhat controversial results.
In particular, Ref.\cite{Kovarik:2010uv} obtains significantly different nuclear PDFs
from fits to charged-lepton and (anti)neutrino DIS data, thus challenging
the QCD factorization theorem \cite{Collins:1989gx}.
However, the analyses of Refs.~\cite{deFlorian:2011fp,Paukkunen:2013grz}
do not support this observation.

Here we follow a different approach and study nPDFs
using the semi-microscopic model developed in Ref.\cite{KP04}.
The model incorporates a number of nuclear corrections including the smearing with the
energy-momentum distribution of bound nucleons (Fermi motion and binding), the off-shell correction to bound
nucleon structure functions, the contributions from meson-exchange currents (MEC), and the propagation of 
the hadronic component of the virtual intermediate boson in the nuclear environment.
The model quantitatively explains the observed 
$x$, $Q^2$ and $A$ dependencies of all the existing nuclear DIS data 
on a wide range of targets from deuteron $^2$H to lead ${}^{207}$Pb~\cite{KP04,KP07,KP10}.

The model of Ref.\cite{KP04} accounts for the modification of PDFs in a bound nucleon
through the off-shell dependence of structure functions. 
In a weakly bound system this effect is described as a linear correction in the nucleon virtuality $p^2-M^2$, with $p$ the nucleon four-momentum and $M$ the nucleon mass \cite{Kulagin:1994fz}. 
The strength of this effect is governed by the relative
response of a PDF to the variation of the nucleon invariant mass $p^2$ in the vicinity
of the mass shell, which is described by a function of the Bjorken variable $x$, $\delta f(x)$. 
We note that, by definition, $\delta f(x)$ describes properties of the nucleon
and in a certain sense can be viewed as a new nucleon structure function.
This function  does not contribute to the cross section of the 
physical nucleon, but it is relevant only for the bound nucleon and describes its 
response to the interaction with the nuclear medium.
The nuclear dependence of this correction is determined by the average nucleon virtuality (off-shellness) in a nucleus.
The off-shell correction proved to be an important contribution to the nuclear  effect 
at large $x$ and was determined phenomenologically from the analysis of data on ratios of 
DIS structure functions in different nuclei \cite{KP04}.  
In a simple single-scale model, in which the quark momentum distributions in the nucleon
are functions of the nucleon radius, the observed behaviour of $\delta f(x)$ can be 
interpreted in terms of the ``swelling'' (i.e., increase of the size)
of the bound nucleon in the nuclear environment. 
In particular, the analysis
of Ref.\cite{KP04} suggests that the nucleon core radius increases by about 10\% in iron, 
while this effect is significantly smaller, about 2\%, in the deuteron.

For simplicity, Refs.~\cite{KP04,KP07,KP10} assume the function $\delta f(x)$ to be universal, i.e. 
flavor independent and also isospin blind (the same for protons and neutrons).
However, our model can naturally incorporate a flavor and isospin dependence into the 
off-shell function, which would then differ for individual parton flavors and types. 
We note that an isospin/flavor dependence of $\delta f$ can lead to predictions similar to other models 
with explicit flavor dependence~\cite{Cloet:2009qs}. 
The study of the flavor and isospin dependence of $\delta f(x)$ requires
nuclear data on high-energy processes which can provide a flavor selection,
like hadronic DY reaction or (anti)neutrino DIS. 
In the present study we use data on DY production to verify the
predictions of the model of Ref.\cite{KP04} and also to address possible differences 
in the off-shell correction between valence and sea-quark distributions.

At small values of $x$ nuclear corrections in DIS are dominated by the effects of 
the propagation of strongly interacting hadronic states in the nuclear environment
\cite{Bauer:iq,Nikolaev:1990ja,Piller:1999wx}.
Such effects can be described in terms of multiple-scattering series
\cite{Glauber:1955qq,Gribov:1968gs}
in the effective scattering amplitude with the relevant quantum numbers.
In this paper we discuss in details how this coherent nuclear correction depends on the
isospin and $C$ parity of the (anti)quark distributions.

We emphasize that the nuclear mechanisms listed above give rise to effects
located in different kinematical regions of Bjorken $x$. 
For instance, the correction related to the nuclear binding (separation) energy is mostly relevant
at large $x \sim 0.5-0.7$ 
\cite{Akulinichev:1985ij,Ku89,Kulagin:1994fz,Kulagin:2008fm}, 
while coherent effects related to the propagation of virtual hadronic states are important
at small $x<0.05$ 
\cite{Bauer:iq,Nikolaev:1990ja,Piller:1999wx}. 
It is important to realize that these effects, which may appear as unrelated,
are actually linked together by the normalization conditions and the energy-momentum sum rules. 
For instance, the momentum sum rule is known to be a useful tool in predicting the anti-shadowing region
in nPDFs~\cite{Nikolaev:1975vy,Frankfurt:1990xz}.
In this paper we use the normalization conditions for the
isoscalar and the isovector valence quark distributions 
as dynamical equations for the effective scattering amplitudes relevant for
the coherent nuclear correction.
These equations are then solved in terms of the off-shell function $\delta f$, 
thus providing a relation between the nuclear shadowing and the off-shell effects. 

Conventionally, we assume that the relevant nuclear constituents are
nucleons interacting via mesonic fields which provide nuclear binding.
The nPDFs are then determined by the convolution of light-cone distribution
function of bound nucleons with the corresponding nucleon PDFs.
The nucleon light-cone distribution functions are driven by the
nuclear spectral function, which defines the energy-momentum distribution of bound nucleons
\cite{Akulinichev:1985ij,Akulinichev:1986gt,Akulinichev:1985xq,Ku89,Kulagin:1994fz,
KP04,KP07,KP10}.
The calculation of mesonic correction is less certain and model-dependent
\cite{LlewellynSmith:1983qa,Ericson:1983um,Berger:1984na,Friman:1983rt,Sapershtein:1985pa,
Kaptari:1989un,Jung:1990pu,Koltun:1997py,Korpa:2013ia}.
However, the nuclear mesonic light-cone distributions are subject to important constrains
from the energy-momentum conservation and from the equations of motion connecting 
the nucleon and the meson correlation functions \cite{Ku89}.
In this paper we further discuss the resulting relations for the moments of
the nuclear meson light-cone distributions
and use them to calculate the mesonic correction to nuclear PDFs.

As an important application of our studies,
we present detailed predictions for the nuclear DY reaction.
The DY process offers a direct probe of the sea quark content in nucleons and nuclei
\cite{McGaughey:1999mq,Peng:2014hta}.
The use of DY data in combination with DIS data allows, thus, a separation of the nuclear corrections for valence  
and sea-quark distributions. The measurements of DY production off nuclear targets by the E772 and E866 
experiments~\cite{E772,E866} at Fermilab do not show any significant enhancement of the sea quark 
distributions in heavy nuclei for $0.1\leq x \leq 0.3$.
Traditionally, this result has been considered in disagreement with the enhancement of the meson cloud 
of a bound nucleon, which is in turn related to the nuclear binding \cite{Bickerstaff:1985ax}.
In this paper we revisit the calculation of nuclear sea- and valence-quark distributions
and argue that the enhancement of nuclear antiquarks owing to the nuclear MEC 
is partially canceled by a negative shadowing correction. 
We examine in detail the ratios of DY cross sections
for different nuclear targets and show that the predicted nPDFs provide a 
good description of both the magnitude and the $x$ and mass dependence of 
the data of Refs.\cite{E772,E866}.

This article is organized as follows. 
In Sec.~\ref{sec:npdf} we review the model of nuclear corrections to PDFs and
discuss their dependence on the type of PDFs. In particular, we study in details 
the nuclear corrections for the isoscalar $q_0=u+d$
and isovector $q_1=u-d$ distributions, 
as well as the dependence of nuclear effects on the $C$ parity of the quark distributions $q^\pm=q\pm \bar q$.
In Sec.~\ref{sec:nDY} we apply our results to the muon pair production
off nuclear targets and provide a detailed comparison of our predictions
with the available data \cite{E772,E866}. In Sec.~\ref{sec:discus} we discuss and summarize our results.

\section{Nuclear Parton Distribution Functions} 
\label{sec:npdf}

It is well known that a PDF describes the momentum distribution of the corresponding parton in a target.
While this is true in a reference frame in which the target has a large momentum (infinite momentum frame), 
the interpretation of PDFs in the target rest frame is somewhat more complicated.
In the target rest frame a PDF also depends on the target energy spectrum 
and it includes the interaction effects of the hadronic component of the virtual 
photon with the target (see, e.g., \cite{Piller:1999wx}).
We recall that in the target rest frame the characteristic propagation time (or longitudinal distance)
of the hadronic fluctuations of the virtual photon is  $L\sim (Mx)^{-1}$, 
where $M$ is the nucleon mass and $x$ the Bjorken scaling variable \cite{Ioffe:1985ep}.
At small $x$, where $L$ is large, diffraction processes dominate the PDFs.
However, when $L$ becomes comparable to the nucleon size, their contribution is reduced.

The scale $L$ can be used to roughtly identify two different kinematical regions for nuclear effects.
At large values of $x$, for which $L<d$, where $d$ is the average distance between bound nucleons, 
nuclear PDFs can be approximated by incoherent contributions from bound protons and neutrons. 
The picture changes at small $x$ ($L\gg d$), where the effects related to the propagation of the 
virtual hadronic (or quark-gluon) states 
in the nuclear medium introduce essential corrections to the impulse approximation.
The interference of multiple scattering contributions and the
energy dependence of the corresponding scattering amplitudes can lead to either a negative (shadowing) 
or a positive (antishadowing) correction, depending on the values of $x$.
It is worth noting 
that this correction, in general, is not universal and may depend on the type of PDF,
as indicated by the studies of Ref.\cite{Kulagin:1998wc,KP04,KP07} and also by phenomenology~\cite{Hirai:2007sx,deFlorian:2011fp,Kovarik:2010uv}.

In this article we study the nuclear quark and antiquark distributions (nPDFs). 
We use the notation $q_{a/T}(x,Q^2)$ for the distribution of quarks of type
$a=u,\ d,\cdots$ in a target $T$. The (anti)quark distribution in a nucleus
receives a number of contributions and can be written as \cite{KP04}
(for brevity, we suppress explicit dependencies on $x$ and $Q^2$)
\begin{equation}
\label{npdf}
q_{a/A} = q_{a/A}^\mathrm{IA} + \delta_{\rm coh}q_a + \delta_\pi q_a,
\end{equation}
where the first term on the right side is the contribution from
bound protons and neutrons in the impulse approximation, and the other
terms are the corrections to the impulse approximation owing to
coherent nuclear interactions of the hadronic component of the virtual photon
and to nuclear meson-exchange currents, respectively.
These contributions are reviewed in the following sections.

\subsection{Impulse approximation and off-shell corrections}
\label{npdf:ia}

It is well known that in the impulse approximation the nPDFs can be written as 
a convolution of the proton (neutron) distribution of a nucleus with the
corresponding PDF of a bound proton (neutron).
The nuclear convolution is an integration over both the nucleon
light-cone momentum $y$ and the nucleon off-shellness (virtuality) $\mu^2$,
since PDFs in an off-shell nucleon generally depend on its virtuality~\cite{Kulagin:1994fz}: 
\begin{equation}
\label{conv:def}
q_{a/A}^\mathrm{IA} = \sum_{\tau=p,n} f_{\tau/A} \otimes q_{a/\tau} = 
\sum_{\tau=p,n} \int_{x<y}\frac{\ud \mu^2\ud y}{y} f_{\tau/A}(y,\mu^2) q_{a/\tau}(\frac xy,Q^2,\mu^2).
\end{equation}
The proton and the neutron distribution function $f$ can be written in terms of the corresponding nuclear
spectral function $\mathcal{P}(\bm{p},\ceps)$ \cite{Akulinichev:1985ij,Akulinichev:1986gt,Ku89,Kulagin:1994fz,KP04}
(for brevity, we drop subscripts identifying the proton and the neutron distributions)
\begin{equation}
\label{D:def}
f(y,\mu^2)= \int [\ud p] \left(1+\frac{p_z}{M}\right)\mathcal{P}(\bm{p},\ceps)
\delta\left(y - \frac{p_0+p_z}{M}\right) \delta\left(\mu^2 - p^2 \right),
\end{equation}
where the integration $[\ud p]=\ud p_0 \ud\bm p/(2\pi)^4$ is performed over the 
nucleon momentum $\bm p$ and energy $p_0=M+\ceps$, and
$p^2=p_0^2-{\bm p}^2$ is the invariant mass of the off-shell nucleon.
We chose a system of coordinates such that the momentum transfer is antiparallel to the $z$ axis.
In the derivation of \eq{conv:def} and \eq{D:def} we also assume the kinematics of the Bjorken limit
and drop any powers of $Q^{-2}$.
Note that the Bjorken variable of the target nucleus is defined in terms of the nucleon mass $M$
and the energy transfer $q_0$ in the target rest frame as $x=Q^2/2Mq_0$.
This variable can vary within the interval $0<x<M_A/M$, where $M_A$ is the mass of a target nucleus.

It should be noted that \eq{D:def} was obtained by expanding a general relativistic expression
in powers of $\bm p/M$ and is valid to the order $\bm p^2/M^2\sim |\ceps|/M$ (including those terms)
\cite{Ku89,Kulagin:1994fz,KP04}.
To this order, the nuclear structure functions in the impulse approximation
are determined by the nonrelativistic nuclear spectral function, which can be written as
\begin{equation}\label{specfn:def}
\mathcal P(\bm p,\ceps) = \int\ud t\, e^{i\ceps t} \average{\psi^\dagger(\bm p,t)\psi(\bm p,0)},
\end{equation}
where $\psi(\bm p,t)$ is the nonrelativistic nucleon operator in the momentum-time representation
(for more details see Ref.\cite{KP04}). By definition, the spectral function describes the energy-momentum
distribution of bound nucleons.
Note that $\ceps$ in \eq{specfn:def} includes the recoil kinetic energy of the residual system of $A-1$ nucleons,
as can be seen after inserting a complete set of states and integrating over the time.
The proton (neutron) spectral function is normalized to the number of bound protons $Z$ (neutrons $N$),
$\int [dp]\mathcal{P}_{p(n)}=Z(N)$.
Using \eq{D:def} we explicitly verify that the proton and neutron distribution functions are normalized
accordingly
\begin{equation}\label{norm}
\int\ud y \ud\mu^2 f_{p(n)/A}(y,\mu^2) = Z (N),
\end{equation}
where the integral is taken over all possible light-cone momenta $y$ and the nucleon virtuality $\mu^2$.
Equations similar to (\ref{npdf}) can be written for the antiquark and
gluon distributions in nuclei.
The distribution functions $f_p(n)$ are independent of $Q^2$ in the Bjorken limit and the $Q^2$ evolution
of nuclear PDFs in the impulse approximation is governed by the evolution
of the PDFs of the corresponding nuclear constituents.
For the discussion of finite $Q$ corrections to the nuclear convolution (\ref{conv:def})
we refer to Ref.\cite{KP04,Ku98}
(see also \cite{Kulagin:2008fm} for spin-dependent DIS).

Note that \eq{conv:def} describes DIS off an off-shell nucleon and for that reason
the bound nucleon PDFs also depend on the invariant mass $p^2$, as an additional variable.
The analysis of the off-shell effect can be significantly simplified by observing that
the nucleon virtuality $v=(p^2-M^2)/M^2$ is, on average, a small parameter \cite{KP04}.
We can then expand the function $q(x,Q^2,p^2)$ as a series
in $v$ in the vicinity of the mass shell $p^2=M^2$, keeping only terms up to the one linear in $v$:
\begin{subequations}
\label{OS:def}
\begin{align}
\label{pdf:os}
q(x,Q^2,p^2) &\approx q(x,Q^2)(1+\delta f(x,Q^2) v),
\\
\delta f(x,Q^2) &= \partial \ln q(x,Q^2,p^2)/\partial\ln p^2 ,
\label{deltaf:def}
\end{align}
\end{subequations}
where $q(x,Q^2)$ is the quark distribution in the on-shell nucleon
and the derivative in \eq{deltaf:def} is evaluated at $p^2=M^2$.
The magnitude  of the off-shell effect is determined by the function $\delta f$.
This function describes the response of the quark distribution in a nucleon to the modification
of its invariant mass owing to interaction effects in the vicinity of the mass shell.
The function $\delta f$ was extracted phenomenologically from an analysis 
of data on the nuclear DIS in Ref.~\cite{KP04}. This analysis suggests
a common off-shell function for the quark and antiquark
distributions, independent of $Q^2$ and of the parton type.  
We further test this assumption in Sec.~\ref{sec:nDY}, by comparing our predictions with data on dimuon
pair production from proton-nucleus collisions. 

Note that, by definition, the function (\ref{deltaf:def}) describes intrinsic properties of the bound nucleon.
The hypothesis that $\delta f$ does not depend on the specific nucleus was verified with a good accuracy
for nuclei ranging from $^{207}$Pb down to $^3$He \cite{KP04,KP10}. We found that the off-shell correction, together
with the nuclear binding correction, is important for a quantitative description of the slope and the position of 
the minimum in the ratio of nuclear structure functions (EMC effect). Overall, this model has been succesfully used to explain
the observed $x$, $Q^2$ and $A$ dependencies of the existing nuclear DIS data on a wide range of targets
from $^2$H to $^{207}$Pb \cite{KP04,KP07,KP10}.

Complex nuclei typically have different numbers of protons and neutrons,
and therefore nuclear PDFs may include both isoscalar and isovector components.
To properly take this effect into account, it is convenient to consider 
the isoscalar $q_0=u+d$ and the isovector $q_1=u-d$ combinations of quark distributions.
Using the isospin symmetry of PDFs, which can be written as
$q_{0/p}=q_{0/n}$ and $q_{1/p}=-q_{1/n}$,
from \eq{npdf} we infer that $q_0$ and $q_1$ are governed by
the isoscalar and the isovector nucleon distributions, respectively,
\begin{subequations}
\label{npdf:01}
\begin{align}
q_{0/A} &= (f_{p/A}+f_{n/A}) \otimes q_{0/p},
\\
q_{1/A} &= (f_{p/A}-f_{n/A}) \otimes q_{1/p},
\end{align}
\end{subequations}
where we use the notations defined in \eq{conv:def}.
It should be emphasized that the separation of the distributions with different isospin
in \eqs{npdf:01} is attributable to the isospin symmetry between $u$ and $d$ quark distributions
in the proton and neutron.%
\footnote{%
Possible violationis of the isospin symmetry in PDFs were discussed in \cite{Londergan:2009kj}.
}
In the present studies we use the isoscalar and the isovector nuclear spectral function of Ref.\cite{KP04}.
Note that the model of spectral function of Ref.\cite{KP04} includes both the mean-field contribution and
a short-range two-nucleon correlation (SRC), giving rise to high-momentum as well as high-energy components
in the spectrum of intermediate nuclear states in \eq{specfn:def}.
We assume the SRC contribution to be similar for the proton and the neutron nuclear distribution.
In particular, we assume that the SRC term only contributes to the isoscalar distribution $f_0=(f_{p/A}+f_{n/A})/A$, 
and that it cancels out in the isovector distribution $f_1=(f_{p/A}-f_{n/A})/A$.
This behavior is supported by
the observation of the dominance of $pn$ SRC pairs in nucleon knock-out experiments \cite{Subedi:2008zz}, 
as well as by a recent analysis of the nuclear momentum distributions in the high-momentum region \cite{Sargsian:2012sm}.
The isovector distribution $f_1$ is calculated as the difference between the mean-field contributions
to the proton and the neutron spectral functions and is proportional to the nuclear asymmetry $\beta=(Z-N)/A$ \cite{KP04}.

The correction driven by the nuclear spectral function (Fermi motion and nuclear binding \cite{Akulinichev:1985ij,Ku89})
along with the off-shell correction \cite{Kulagin:1994fz} are the leading nuclear effects at large $x$, as verified
by the extensive studies of Refs.\cite{KP04,KP10}.
At small $x$ there are significant corrections to the impulse approximation.
It is worth mentioning that to satisfy the nuclear light-cone momentum sum rule, 
contributions from degrees of freedom other than nucleons are required.  
Indeed, we can obtain the fraction of the nuclear light-cone momentum carried by nucleons after integrating
the nucleon distribution function,
\begin{equation}\label{y:ia}
\average{y}_N=\int \ud y\ud p^2 f_0(y,p^2) y=  1+\frac{\average{\ceps} +\tfrac23\average{T}}{M} ,
\end{equation}
where  $\average{\ceps}$ and $\average{T}=\average{\bm p^2}/2M$
are the nucleon separation and kinetic energy, respectively, averaged with the isoscalar nuclear spectral function.%
\footnote{%
Note that our definition of $\ceps$ includes the energy of the recoil nucleus
such that, e.g., for the deuteron $\ceps=\ceps_D-{\bm p}^2/2M$, where $\ceps_D\approx -2.2\ \Mev$
is the the deuteron binding energy.
See Refs.~\cite{KP10,Kulagin:2008fm} for a discussion about the relation between $\ceps$ and the ``conventional"
separation energies and spectral functions.
}
The correction to unity in \eq{y:ia} is negative, suggesting that the impulse approximation violates 
the nuclear light-cone momentum balance.
This is not unexpected because the fields responsible for the nuclear binding 
also carry the missing light-cone momentum and therefore they should be considered explicitly.

\subsection{Correction owing to nuclear meson-exchange currents}
\label{nuclear:pi}

The correction originated from the virtual mesons exchanged between bound nucleons 
was extensively discussed in the context of the nuclear EMC effect 
\cite{LlewellynSmith:1983qa,Ericson:1983um,Berger:1984na,Friman:1983rt,Sapershtein:1985pa,Jung:1990pu}.
Following the approach of Sec.\ref{npdf:ia}, 
this correction can be written in terms of the convolution (\ref{conv:def}) of the nuclear pion 
distribution function with the (anti)quark distribution in a virtual pion.
Pions can be in three possible charge states: $\pi^0, \pi^+, \pi^-$. 
Similarly to Sec.\ref{npdf:ia}, we separate the pion corrections for the isoscalar and the isovector nuclear PDFs.
Assuming the isospin symmetry of the quark distributions in different pion states,
$q_{0/\pi^+}=q_{0/\pi^-}=q_{0/\pi^0}$ and $q_{1/\pi^+}=-q_{1/\pi^-}$ and $q_{1/\pi^0}=0$, we have \cite{KP04}
\begin{subequations}
\label{npdf:pi}
\begin{align}
\delta_\pi q_{0/A}(x,Q^2) &= f_{\pi/A} \otimes q_{0/\pi},\\
\delta_\pi q_{1/A}(x,Q^2) &= (f_{\pi^+/A}-f_{\pi^-/A}) \otimes q_{1/\pi^+}.
\end{align}
\end{subequations}
The pion distribution entering in the first equation, $f_{\pi/A}$, is the sum
over the pion states $\pi^+,\pi^0$, and $\pi^-$. The pion distributions refer only to the nuclear pion excess,
since the scattering off virtual pion emitted and absorbed by the same nucleon (nucleon pion cloud) 
should be already included into the proton and neutron PDFs.

The isospin symmetry suggests an equal distribution of pion quarks and antiquarks in the isoscalar combination,
thus $q_{0/\pi}^-=0$. For this reason, the pion correction to the isoscalar nuclear valence quark distributions vanishes,
$\delta_\pi q_{0/A}^-=0$.  
However, the nuclear sea is obviously affected by the pion contribution.
Note that the isovector part of the valence quark distribution $q_{1/\pi}^-$ is finite.
Thus nuclear pions, in general, do contribute to the isovector nuclear valence distribution $q_{1/A}^-$.
This correction is driven by the $\pi^+ -\pi^-$
asymmetry of the nuclear pion distributions, as it can be seen from \eq{npdf:pi}.
In the present discussion we assume for simplicity identical $\pi^+$ and $\pi^-$ nuclear distributions
and postpone the discussion of the isovector pion effect for future studies.
Therefore, we also have $\delta_\pi q_{1/A}^-=0$.

The pion light-cone distribution can be written as \cite{Ku89,KP04}
\begin{eqnarray}
f_{\pi/A}(y,\mu^2) &=& 2yM \int[\ud k]\mathcal{D}_{\pi/A}(k)
                \delta\left(y-\frac{k_0+k_z}{M}\right)\delta(\mu^2-k^2),
\label{pion:f}
\\
\mathcal{D}_{\pi/A}(k) &=& \int \ud t \exp(ik_0 t)
        \average{\varphi^\dagger(\bm{k},t)\varphi(\bm{k},0) },
\label{pion:D}
\end{eqnarray}
where, similarly to \eq{specfn:def}, the averaging is taken over the nuclear ground state
and $\varphi$ is the pion field operator in the momentum-time space.%
\footnote{$\varphi(\bm k,t)$ is the Fourier transform of the pion field operator $\varphi(\bm r,t)$
in the Heisenberg representation.}
We also assume the sum over different pion isospin states.

The pion energy-momentum distribution $\mathcal{D}_{\pi/A}(k)$ is proportional to
the imaginary part of the full pion propagator in a nucleus,
which is, in turn, proportional to the spin-isospin nuclear response function.
This relation was exploited in a number of calculations of the nuclear
pion/meson correction to the nuclear structure functions
\cite{Ericson:1983um,Sapershtein:1985pa,Jung:1990pu,Korpa:2013ia}. 
These studies show an enhancement of the nuclear structure functions in the region $0.05<x<0.15$.
However, it should be noted that the specific results on the nuclear pion distribution
are sensitive to the details of the pion-nucleon form-factor, as well as to the treatment
of the particle-hole nuclear excitations
(i.e. uncertainties in the values and possible energy-momentum dependence of the Landau-Migdal parameters),
and of the $\Delta$ degrees of freedom in the response function.

We use a different approach \cite{Ku89}.
In the following we discuss the constraints on the pion distribution function, which can be derived
by imposing the nuclear light-cone balance and by studying the meson contribution to
the nuclear potential energy.
We focus on the dependence on the light-cone variable and
do not discuss the off-shell effect on the virtual pion PDFs.
To this end, we integrate the distribution (\ref{pion:f})
over the pion virtuality, $f_{\pi/A}(y)=\int \ud \mu^2 f_{\pi/A}(y,\mu^2)$.

Equation~(\ref{pion:f}) defines an antisymmetric function $f_{\pi/A} (-y) = -f_{\pi/A} (y)$.
This property allows us to derive the sum rules for the odd moments of the pion distribution
function in the physical region of $y>0$.
We notice that  the first moment of (\ref{pion:f})
reduces to the light-cone component of the pion energy-momentum tensor 
$\theta^\pi_{++}=\left(\partial_0\varphi\right)^2 +\left(\partial_z\varphi\right)^2$
in a nucleus  
\begin{equation}
\label{y:pi}
\average{y}_\pi =\int\ud y yf_{\pi/A}(y)=\average{\theta_{++}^\pi}/M.
\end{equation}
It was shown in Ref.\cite{Ku89} 
that the nucleon and the pion distribution functions
[Eqs.~(\ref{D:def}) and (\ref{pion:f})] are consistent with 
the light-cone momentum balance equation
\begin{equation}\label{balance:eq}
\average{y}_\pi + \average{y}_N = M_A/(AM),
\end{equation}
where $M_A=A(M+\ceps_B)$ is the nucleus mass and $\ceps_B$ is the nuclear binding energy per nucleon.

To further constrain the pion distribution (\ref{pion:f}), 
we consider the average $y^{-1}$, which is proportional 
to $\varphi^2$ averaged over the nuclear ground state \cite{Ku89},
\begin{equation}
\label{invy:pi}
\average{y^{-1}}_\pi =\int{\ud y} y^{-1}f_{\pi/A}(y) =M\average{\varphi^2}.
\end{equation}

A number of constraints on the nuclear pion distribution
$\mathcal{D}_{\pi/A}(k)$ can be obtained in a model with a nuclear Hamiltonian
including nucleons interacting with the pion field.
Consider the pion kinetic term in the nuclear Hamiltonian.
Its mean value over the nuclear ground state can be written as
\begin{equation}
\label{pion:energ}
\average{(\nabla\varphi)^2 + m_\pi^2\varphi^2} = -\average{V_\pi},
\end{equation}
where $\average{V_\pi}$ is the contribution to the nuclear potential energy
owing to the one-pion exchange potential, averaged over the nuclear ground state.
This relation can be derived by using the equation of motion for the pion field operator in the
static approximation $\partial_0\varphi=0$ \cite{Ku89}.
In this context we note that the pion field in nuclei is generated by
nucleon sources and its time dependence describes retardation
effects in the nucleon--nucleon interaction. In a nonrelativistic system
this effect is small because typical energy variations are small compared to
the pion mass.
In the same approximation, for the pion energy-momentum tensor we have 
$\average{\theta_{++}^\pi}=\frac13\average{(\nabla\varphi)^2}$.
Using this relation in \Eqs{y:pi}{pion:energ}, we obtain
\begin{subequations}
\label{av:pion}
\begin{align}
\label{av:nablaphisq}
\average{(\nabla\varphi)^2} &= 3M\average{y}_\pi ,
\\
\label{av:phisq}
m_\pi^2\average{\varphi^2} &= -\average{V_\pi} - 3M\average{y}_\pi .
\end{align}
\end{subequations}

It is interesting to note that the normalization of the pion distribution (\ref{pion:f}),
i.e. the average pion excess number in a nucleus, can be constrained in terms of the moments
$\average{y}_\pi$ and $\average{y^{-1}}_\pi$. Indeed, assuming $f_{\pi/A}\ge 0$,
we apply the triangle inequality to the distribution (\ref{pion:f}) and obtain
\begin{equation}\label{pion:N}
N_\pi=\int \ud y f_{\pi/A}(y) < 
\left( \average{y}_\pi \average{y^{-1}}_\pi \right)^{1/2}.
\end{equation}

At this point it should be noted that the pion exchange alone is not sufficient to describe the nucleon-nucleon
interaction. Other mesons, such as scalar $\sigma$, vector $\omega$ and $\rho$, contribute
to both the nucleon-nucleon interaction and the nuclear DIS.
Their contribution to the nPDFs is described by
equations similar to \Eqs{pion:f}{pion:D}, with the pion field replaced by the corresponding mesonic field.
The light-cone balance equation (\ref{balance:eq}) and the pion contribution to the nuclear potential
energy in \eq{pion:energ} can be generalized to include other mesonic contributions.
Let us consider the light-cone distribution function corresponding
to the sum of $\pi,\ \rho,\ \omega$ and $\sigma$ mesons:
\begin{equation}
\label{mes:f}
f_M(y)=\sum_{m=\pi,\rho,\ldots} f_m(y)
\end{equation}
The moments of this distribution can be written similarly to \Eqs{y:pi}{invy:pi} as 
\begin{align}
\label{y:M}
\average{y}_M &=\frac{1}{3M}\sum_m \average{(\nabla\phi_m)^2},
\\
\label{invy:M}
\average{y^{-1}}_M &=M\sum_m \average{\phi_m^2},
\end{align}
where $\phi_m^2$ is the corresponding meson field squared for the (pseudo)scalar mesons, 
and we have $\phi_\omega^2=\bm\omega^2-\omega_0^2$ for the vector mesons and a similar
term for the $\rho$ mesons. 

The generalization of the light-cone balance equation is straightforward,
because \eq{balance:eq} holds in the presence of several meson fields, with 
$\average{y_\pi}$ replaced with the total meson light-cone momentum  $\average{y_M}$.
The pion energy equation (\ref{pion:energ}) can also be generalized for the 
presence of several meson fields
\begin{equation}
\label{mes:energ}
\sum_m \average{(\nabla\phi_m)^2 + m_m^2\phi_m^2} = -\average{V},
\end{equation}
where $m_m$ is the mass of the corresponding meson.
In the right-hand side of \eq{mes:energ} the term $\average{V}$ is the nuclear potential energy,
i.e., the full one-meson-exchange potential $V=V_\pi + V_\omega + V_\rho + V_\sigma$
averaged over the nuclear ground state. The nuclear potential energy
$\average{V}$ is related to the mean separation and kinetic energy as (see Sec.3 of Ref.\cite{Ku89} for more detail)
\begin{equation}\label{av:V}
\average{\ceps} = \average{T} + \average{V} .
\end{equation}
Using \Eqs{y:M}{mes:energ} we can estimate the average $\phi^2_m$, which  determines the moment $\average{y^{-1}}_M$,
\begin{equation}
\label{av:mes}
m_M^2 \sum_m \average{\phi_m^2} = -\average{V} -3M\average{y}_M,
\end{equation}
where $m_M$ is an average meson mass.

We use the constraints and equations discussed above to model the nuclear meson distribution (\ref{pion:f}).
It is important to note that \eqs{y:M} through (\ref{av:mes}) allow us to constrain the overall behavior of
the meson distribution in terms of the nucleon spectral function (\ref{specfn:def}), and the
energy parameters $\average{\ceps}$ and $\average{T}$.
We must consider a realistic parametrization of the distribution (\ref{pion:f}).
We first note that Eq.(\ref{pion:f}) shows a linear dependence on $y$ as $y\to0$.
The distribution (\ref{pion:f}) is concentrated mainly in the region $y \sim k_M/M$, 
where $k_M$ is a typical virtual meson momentum, which can be estimated as
\begin{equation}
\label{mes:k}
k_M^2 = \sum_m \average{(\nabla\varphi_m)^2}/\sum_m \average{\varphi_m^2}=
3M^2\average{y}_M/\average{y^{-1}}_M .
\end{equation}
Note that this equation gives the average pion momentum in terms of the moments
$\average{y}_M$ and $\average{y^{-1}}_M$ of the light-cone distribution (\ref{mes:f}).
Using \eqs{y:M} through (\ref{av:mes}) and $\average{\ceps}$ and $\average{T}$
calculated with the spectral function of Ref.\cite{KP04},  
we obtain $\average{y}_M=0.023$ and $\average{y^{-1}}_M=0.954$ for iron, 
while for the deuteron we obtain $0.0045$ and $0.402$, respectively.
These values suggest that the characteristic value of $y\sim k_M/M$ spans the region  
$0.2-0.3$ for light and heavy nuclei.
The region $y\sim 1$ requires relativistic momenta of virtual mesons $\sim 1\ \gev$, which we assume to
be suppressed. On the basis of these arguments, we consider the following  model for the meson
distribution in the region $0<y<1$
\begin{equation}
\label{fpi:model}
f_{M/A}(y) =c\,y(1-y)^n.
\end{equation}
The parameters $c$ and $n$ are fixed from $\average{y}_M$ and $\average{y^{-1}}_M$.
By integrating \eq{fpi:model}, we obtain an average meson number 
$N_M=0.11$ for the iron nucleus and $0.031$ for the deuteron.

\subsection{Effects owing to propagation of intermediate hadronic states}
\label{npdf:coh}

In this section we review corrections arising 
owing to the propagation of the hadronic component of the intermediate boson in the nucleus rest frame.
These effects are relevant at low values of the Bjorken $x$,
as the virtual hadronic states have an average lifetime (or the correlation length) $L\sim (Mx)^{-1}$ \cite{Ioffe:1985ep}.
For the leading contribution to the DIS structure functions
the coherent multiple scattering interactions of the intermediate states with the nucleons 
lead to a negative correction known as the nuclear shadowing effect,
as discussed in a number of studies (for a review, see Ref.\cite{Piller:1999wx}).

We follow the approach developed earlier in Refs.\cite{KP04,KP07} for the structure functions, 
and evaluate the corresponding corrections to nuclear PDFs.
We will approximate the sum over the set of intermediate hadronic states by a single
effective state and describe its interaction with the nucleon with an effective scattering amplitude $a$.
Let $a_{qp}$ and $a_{\bar qp}$ be the effective proton scattering amplitude of this state
corresponding to the quark and antiquark distributions of type $q$ in the proton.
Similar notations are used for the neutron distributions.
The ratio $\delta\mathcal R=\delta_\mathrm{coh} q_N/q^\mathrm{coh}_N$ describes
the relative nuclear effect in the coherent component of the quark distribution.
Using the optical theorem this ratio can be written in terms of effective cross sections, 
or the imaginary part of the effective amplitudes in the forward direction
\begin{equation}
\delta \mathcal{R}_f = \Im \mathcal T^A(a_f)/(A\,\Im a_f)
\end{equation}
where $\mathcal T^A(a)$ is the sum of the nuclear multiple-scattering series 
driven by the propagation of the intermediate hadronic states in a nucleus.
Note that the multiple-scattering series should start from the double-scattering term, 
as the single-scattering term is already accounted in the impulse approximation of \eq{npdf}.
The elastic scattering amplitude $a(s,k)$ depends on the center-of-mass energy $s$ and the momentum transfer $k$.
We choose a normalization of the amplitude such that the optical theorem reads $\Im a(s,0)=\sigma(s)/2$,
where $\sigma$ is the total cross section, and write the amplitude as $a=(i+\alpha)(\sigma/2)\exp(-B k^2/2)$,
where the exponent describes the dependence on the momentum transfer.

For the deuteron we only have a double scattering term and the amplitude $\mathcal T^D$ can be written as
\begin{align}
\label{TD}
\mathcal T_D &= i a_p(0) a_n(0)\int\frac{\ud^2k_\perp}{(2\pi)^2} S_D(k_\perp,k_z) e^{-B(k_\perp^2+k_z^2)} ,
\\
\label{SD}
S_D(k) &= \int\ud^3 r e^{i\bm k\cdot\bm r} |\Psi_D(\bm r)|^2,
\end{align}
where $\Psi_D$ is the deuteron wave function,
$S_D(k)$ is the deuteron elastic form factor,
and $a_{p(n)}(0)$ denotes the proton (neutron)
scattering amplitude in the forward direction.
Equation (\ref{TD}) is similar to the well-known Glauber formula \cite{Glauber:1955qq}.
However, we should note the dependence on the longitudinal momentum transfer $k_z$,
which is not present in \cite{Glauber:1955qq}.
The longitudinal momentum transfer $k_z$ develops because of inelastic transitions
and depends on the mass of the states produced diffractively \cite{Gribov:1968gs}.
We set $k_z=Mx$ to account for a finite longitudinal correlation length
of the hadronic component of the intermediate boson at high $Q^2$.

For heavy nuclei the double scattering term has a form similar to \eq{TD} in the optical approximation 
\begin{align}
\label{TA2}
\mathcal T_A^{(2)}(a) &= \frac{i}{2}(1-{A}^{-1}) \int\frac{\ud^2k_\perp}{(2\pi)^2} a(k)S(k) a(-k)S(-k),
\\
S(k) &= \int \ud^3\bm r \exp(i\bm k\cdot\bm r)\rho(\bm r),
\end{align}
where $\rho$ is the nuclear density (normalized to the number of particles)
and  $aS=a_p S_p+a_n S_n$ is the sum of the proton and the neutron terms
with the corresponding density distributions. 
We note that \eq{TA2} holds for a generic nucleus of $Z$ protons and $N$ neutrons. Therefore, 
it accounts for both isoscalar and isovector effects in the nuclear shadowing correction.
We discuss the separation of these effects later in this section.

For heavy nuclei the multiple-scatterring series goes beyond the double-scattering term 
[although the double-scattering correction dominates if the mean free path of the hadronic state is larger
than the nuclear radius, $(\rho\sigma)^{-1} > R_A$].
The sum of the Glauber multiple-scattering series 
can be written in a compact form for a $A\gg 1$ nucleus in the optical approximation
(see, e.g., Ref.\cite{Bauer:iq})
\begin{align}\label{TA}
\mathcal T^A(a) &= i \int_{z_1<z_2}\hspace{-1.5em}
\ud^2\bm{b}\ud z_1\ud z_2\,a \rho^B(\bm{b},z_1) a \rho^B(\bm{b},z_2)
    e^{i\int_{z_1}^{z_2}\ud z' a \rho^B(\bm{b},z')}
    e^{ik_z(z_1-z_2) - B k_z^2},
\end{align}
where the integration is performed along the collision axis, 
chosen to be the $z$ axis, and over the transverse positions of 
the nucleons (impact parameter $\bm{b}$),
$a \rho^B=a_p(0)\rho_p^B+a_n(0)\rho_n^B$ with $\rho_p^B$ and $\rho_n^B$ the proton and the neutron
density convoluted with the profile function of the scattering amplitude in the impact parameter space:
\begin{equation}
\label{rhoB}
\rho^B(\bm b,z) = \int\ud^2\bm b' \frac{ \exp(-\frac{(\bm b-\bm b')^2}{2B}) }{2\pi B} \rho(\bm b',z)
\end{equation}
In \Eqs{TA}{rhoB} we use the $\exp(-Bk^2/2)$ momentum transfer dependence of the scattering amplitude. 
Note that the proton and the neutron densities are normalized to the proton $(Z)$ and the neutron $(N)$ numbers, 
respectively. 
The density in the exponential factor of \eq{TA} accounts for multiple scattering 
effects (i.e. triple and higher order rescattering).
Equation(\ref{TA}) reduces to \eq{TA2} in the double scattering approximation, up to a $1/A$ term.

We now separate the isoscalar and the isovector contributions in \eq{TA},
relevant for the $u$ and $d$ quark distributions.
To this end, we assume the isospin symmetry for the scattering off protons and neutrons, i.e.
$a_{up}=a_{dn}$ and $a_{dp}=a_{un}$, and write the amplitudes as 
$a_{up}=a_0+\tfrac12 a_1$ and $a_{dp}=a_0-\tfrac12 a_1$, 
where $a_0$ and $a_1$ are the isoscalar and isovector amplitudes, respectively.
To the first order in $\beta=(Z-N)/A$ we have \cite{KP07}
\begin{equation}\label{TA:ud}
\mathcal{T}^A(a_{u,d}) = \mathcal{T}^A(a_0)\pm \tfrac12{\beta} a_1\mathcal{T}^A_1(a_0),
\end{equation}
where $+$ should be taken for $u$ quark, and $-$ for $d$ quark, 
and $\mathcal T^A_1(a)=\partial\mathcal T^A(a)/\partial a$.
The first and the second terms in  \eq{TA:ud} determine the corrections to the 
isoscalar and the isovector quark distributions, respectively.

An equation similar to \eq{TA:ud} can be obtained for the antiquark amplitudes. 
It is convenient to discuss combinations of PDFs with definite isospin and $C$ parity.
We define  $q_I^\pm = q_I \pm \bar q_I$ with $I=0,1$ and 
combine the quark and antiquark amplitudes  to derive the  
nuclear corrections to these PDFs in terms of the ratios 
$\smash{\delta\mathcal R_I^C=\delta_{\rm coh}q_I^C/(A\,q^C_{I/p})}$
(note that $\smash{q_{I/p}^C}$ are the proton PDFs).
We first consider the isoscalar $I=0$ case.
For coherent nuclear corrections to the $C$-even and $C$-odd quark distributions we have 
\begin{subequations}\label{npdf:coh:0}
\begin{align}
\label{coh:0pl}
\delta \mathcal{R}^{+}_0 &= \Im \mathcal T^A(a_0^+)/(A\,\Im a_0^+) ,
\\
\label{coh:0mn}
\delta \mathcal{R}^{-}_0 &= \Im [a_0^{-} \mathcal T^A_1(a_0^{+})]/(A\, \Im a_0^{-}) ,
\end{align}
\end{subequations}
where $a_0^\pm = a_0 \pm \bar a_{0}$ are the isospin 0 amplitudes with $C$ parity $C=\pm 1$. 
We note that \eqs{npdf:coh:0} were obtained by treating the $C$-odd amplitude as a small parameter and
expanding the difference between the quark and antiquark nuclear 
amplitudes in series of $a_0^-$ to the order $(a_0^-)^2$ \cite{KP07}.
The effective expansion parameter in \eqs{npdf:coh:0} is the ratio of the amplitudes
$a_0^-/a_0^+$. The smallness of this parameter can be justified
within the Regge pole model of high-energy scattering amplitudes. 
Indeed, the Pomeron gives the leading contribution to the $C$-even amplitude $a_0^+$.
However, its contribution cancels out in the $C$-odd amplitude $a_0^-$, which is determined
by subleading Regge poles.

It should be noted that the $C$-odd ratio $\mathcal R^-_0$ from \eq{coh:0mn}  
is independent of the $C$-odd cross section $\sigma_0^-$,
but depends on the ratio $\alpha_0^- = \Re a_0^-/\Im a_0^-$ and on the $C$-even cross section, 
which determines the rate of nuclear effects on parton distributions.
The result is also affected by the interference of the real parts of the amplitudes in the
$C$-even and $C$-odd channels.
It is interesting to note that we obtain a simple relation between $\mathcal R^+_0$ and $\mathcal R^-_0$
if we only consider the double scattering term in \eqs{npdf:coh:0}.
We have  \cite{KP04}
\begin{equation}
\label{r:mn:pl}
\frac{\delta\mathcal R^-_0}{\delta \mathcal R^+_0} = 2\,\frac{1-\alpha_0^- \alpha_0^+}{1-{\alpha_0^+}^2}
\end{equation}
This equation suggests that the relative nuclear effect for the $C$-even and the $C$-odd cross sections
is independent of the cross section and only depends on the $\Re/\Im$ ratios of the amplitudes.%
\footnote{%
\eq{r:mn:pl} holds at small $x$, such that the phase $\exp[ik_z(z_1-z_2)]$ in \eq{TA} can be neglected.
}
In case of vanishing $\alpha_0^+$ the relative $C$-odd shadowing effect is enhanced by a factor of 2
\cite{Kulagin:1998wc}.

We now discuss the isovector coherent (shadowing) correction to the nuclear (anti)quark distributions.
To this end, we consisder the multiple scattering corrections to the $C$-even and $C$-odd 
isovector combination $a_u-a_d$ using \eq{TA:ud}.
Similarly to the isoscalar case discussed above, we expand the terms $\mathcal T_1^A(a_0^+\pm\tfrac12 a_0^-)$
in \eq{TA:ud} in series of $a_0^-$. To first order we obtain 
\begin{subequations}\label{npdf:coh:1}
\begin{align}
\label{coh:1pl}
\delta \mathcal{R}^{+}_1 &= 
		\beta \Im \left[
	a_1^{+} \mathcal T^A_1(a_0^+) + \tfrac14 a_1^- a_0^- \mathcal T_2^A(a_0^+)
\right]/ (A\,\Im a_1^+) ,
\\
\label{coh:1mn}
\delta \mathcal{R}^{-}_1 &= 
		\beta \Im \left[
	a_1^{-} \mathcal T^A_1(a_0^{+}) + a_1^+ a_0^- \mathcal T_2^A(a_0^+)
\right]/(A\, \Im a_1^{-}) ,
\end{align}
\end{subequations}
where $\mathcal T_2(a)=\partial \mathcal T_1(a)/\partial a$.

The corresponding individual corrections for $u$ and $d$ quarks and antiquarks
are given in terms of the isoscalar $(q_{0/p}=u+d)$ and the isovector $(q_{1/p}=u-d)$
components of the quark distributions in the proton and
 $\delta\mathcal R_{0,1}^\pm$ as
\begin{subequations}\label{coh:qqbar}
\begin{align}
\label{coh:ud}
\delta\mathcal{R}_{u,d} &= \delta\mathcal{R}^{+}_0 +
	\frac{q_0^-}{2q_0}\left( \delta\mathcal{R}^{-}_0 - \delta\mathcal{R}^{+}_0 \right)
	\pm \left[
		\frac{q_1}{q_0}\delta\mathcal{R}^{+}_1 
			+ \frac{q_1^-}{2q_0}\left( \delta\mathcal{R}^{-}_1 - \delta\mathcal{R}^{+}_1 \right)
	    \right],
\\
\label{coh:udbar}
\delta\mathcal{R}_{\bar u,\bar d} &= \delta\mathcal{R}^{+}_0 -
	\frac{q_0^-}{2\bar q_0}\left( \delta\mathcal{R}^{-}_0 - \delta\mathcal{R}^{+}_0 \right)
	\pm \left[
		\frac{\bar q_1}{\bar q_0}\delta\mathcal{R}^{+}_1 
			- \frac{q_1^-}{2\bar q_0}\left( \delta\mathcal{R}^{-}_1 - \delta\mathcal{R}^{+}_1 \right)
	    \right],
\end{align}
\end{subequations}
where the sign $+$ should be taken for $u$ quarks, and the sign $-$ for $d$ quarks. 
We recall that $q_{0,1}=u \pm d$, $\bar q_{0,1}=\bar u \pm \bar d$, and $q_{0,1}^-=u_{\rm val}\pm d_{\rm val}$ 
are the (anti)quark distributions for the proton taken for the given $x$ and $Q^2$.

The effective amplitudes $a$ with either isospin 1 or $C=-1$ are generally significantly smaller
than the leading amplitude $a_0^+$, which  drives multiple scattering corrections
for all distributions, as can be seen from \Eqs{npdf:coh:0}{npdf:coh:1}.
If only  linear terms in $a_0^-$ and $a_1^\pm$ are retained, then the corresponding nuclear
ratios depend on the $\alpha=\Re a/\Im a$ ratios of these amplitudes.

\subsection{Normalization constraints}
\label{sec:norm}

The PDFs obey a number of sum rules reflecting the general symmetries of the strong interaction.
Important examples include the valence quark number sum rule,
for both the isoscalar and the isovector channels,
and the light-cone momentum sum rule. 
Because of the underlying  symmetries,
these sum rules should not be affected by the strong interaction, including the nuclear effects. 
Therefore, for any particular model it is important to explicitly verify that a cancellation
of different nuclear effects occurs in the PDFs sum rules.

We first consider the sum rule of the isoscalar valence quark number per bound nucleon:
\begin{equation}\label{qval:sr}
N_\mathrm{val}^A = A^{-1} \int_0^A\ud x\,q_{0/A}^- = 3,
\end{equation}
where $q_0^-=u^-+d^-$ is the isoscalar valence quark distribution%
\footnote{%
We do not consider the $s^-$ and $c^-$ quark distributions. In general $s^-(x)\not=0$,
but this gives vanishing contribution to \eq{qval:sr}.
}.
We consider now the contributions to \eq{qval:sr} from the various nuclear effects present in our model.
First we explicitly calculate the normalization in the impulse approximation by \eq{conv:def} and obtain
\begin{align}
N_\mathrm{val}^{\rm IA} &= 3 + \delta N_{\mathrm{val}}^{\mathrm{OS}},
\\
\label{del:val:os}
\delta N_{\mathrm{val}}^{\mathrm{OS}} &= \average{v}_0\int_0^1 \ud x\, q_{0/N}^-(x) \delta f(x), 
\end{align}
where $N_\mathrm{val}^N=3$ is the valence quark number in the nucleon,
$\average{v}_0=\average{p^2{-}M^2}/M^2$ is the nucleon virtuality 
averaged with the nuclear spectral function (the subscript 0 indicates that we should take the isoscalar part),
and $\delta f$ is the off-shell function defined in \eq{deltaf:def}.
Note that in the absence of the off-shell correction ($\delta f=0$) \eq{qval:sr} has already the 
correct normalization because 
nuclear effects owing to the nuclear spectral function cancel out in the valence quark normalization. 
The off-shell (OS) correction, in general, does not vanish.
As discussed in Sec.~\ref{npdf:ia}, we assume a universal flavor-independent OS function $\delta f(x)$,
common to quark and antiquark distributions. This assumption is supported by the analysis of Ref.\cite{KP04},  
which allowed a precise determination of this correction 
from the measured ratios of structure functions in nuclear DIS.  
In Sec.~\ref{sec:nDY} we further verify the universality of $\delta f$ for all partons 
by studying the nuclear DY process.

The nuclear meson correction to the nuclear valence distribution cancels out (Sec.~\ref{nuclear:pi}).
However, the nuclear coherent (coh) effects give a nonzero contribution to the valence quark normalization,
\begin{equation}
\label{del:val:sh}
\delta N_{\mathrm{val}}^\text{coh} = \int_0^1\ud x\,q_{0/N}^-(x)\delta \mathcal{R}_{0}^- ,
\end{equation}
where $\delta \mathcal{R}_{0}^-$ is given by \eq{coh:0mn}.
To satisfy \eq{qval:sr} we require a cancellation
between the OS and the coh corrections in the valence quark normalization,
\begin{equation}
\label{nbalance}
\delta N_{\text{val}}^{\text{OS}}+\delta N_{\text{val}}^{\text{coh}}=0.
\end{equation}

\begin{figure}[htb]
\centering
\includegraphics[width=0.85\textwidth]{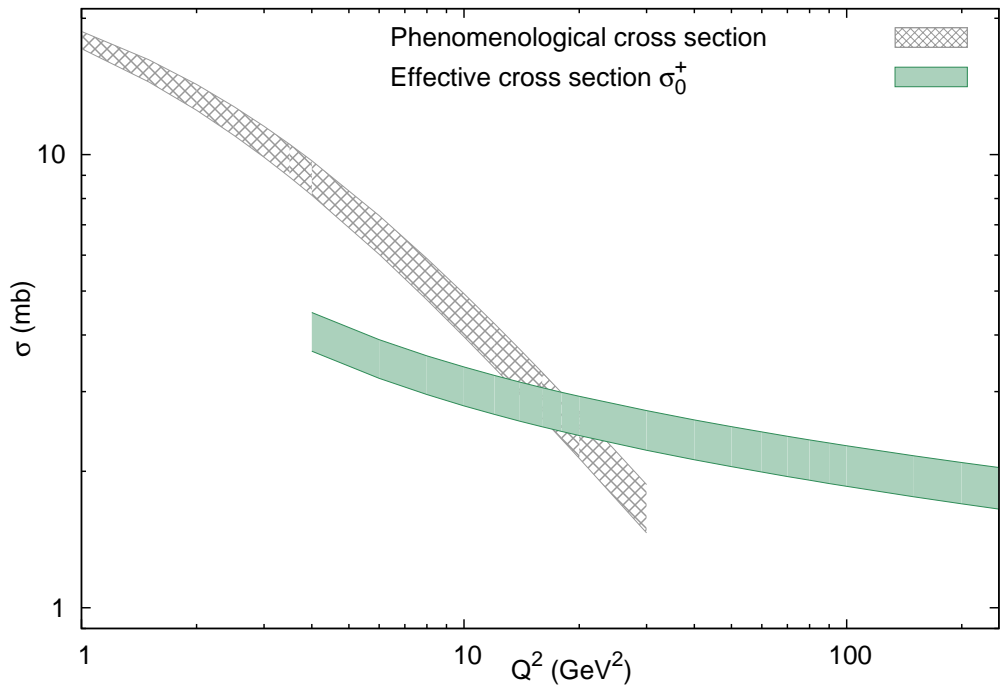}
\caption{%
(Color online)
Effective cross section $\sigma_0^+$ calculated from the normalization condition (\ref{nbalance}) (see text) 
as a function of $Q^2$. The phenomenological cross section of Ref.~\cite{KP04} is also shown for comparison. 
\label{fig:xsec}}
\end{figure}

It is worth noting that the nuclear (anti)shadowing correction is an effect related to small $x$ values,
while the OS correction is mainly located at large $x$.
Therefore, the normalization constraint introduced by \eq{nbalance} provides a nontrivial connection
between nuclear effects of completely different origin.  
In the present analysis we use the off-shell function $\delta f(x)$ of Ref.\cite{KP04}
to calculate the OS correction to the normalization $\delta N_{\text{val}}^{\text{OS}}$.
We then use \eq{nbalance} in order to constrain the effective amplitudes 
$a_0^+$ and $a_0^-$ in the region of high $Q^2$.
To this end, we note that in \eq{coh:0mn} the relevant correction $\delta\mathcal{R}_{0}^-$ depends on the
$C$-even cross section $\sigma_0^+=2\Im a_0^+$ and the phases $\alpha=\Re a/\Im a$, both $C$-even and $C$-odd,
responsible for the interference effects in the multiple scattering series.
For simplicity we assume that the effective cross section $\sigma_0^+$
and the phases $\alpha_i^c$ are independent of energy
in the high-energy region, corresponding to small $x$, 
and we fix $\alpha_0^+=-0.2$ using the results of Ref.\cite{KP04}.
From \eq{nbalance} we calculate $\sigma_0^+(Q^2,\alpha)$ as a function of $Q^2$ and the 
$C$-odd phase $\alpha=\alpha_0^-$. Note that the phase $\alpha_0^-$ is not directly constrained by \eq{nbalance}. 
We determine this parameter by requiring $\sigma_0^+(Q^2,\alpha)$
to match the corresponding phenomenological cross section of Ref.\cite{KP04} in the region of 
$15 \lesssim Q^2\lesssim 20\ \gevsq$.%
\footnote{This choice is motivated by the fact that we need values of $Q^2$ which are  
sufficiently large to suppress higher-twist contributions, but at the same time which are still in a 
kinematic region constrained by the available data on nuclear shadowing.}
We obtain $\alpha_0^-=1.41$ 
and the cross section $\sigma_0^+$ shown in Fig.~\ref{fig:xsec}, 
together with the phenomenological cross section of Ref.\cite{KP04} calculated from the 
analysis of nuclear shadowing data with $Q^2\lesssim 20\ \gevsq$.
The $1\sigma$ error band for the effective cross-section $\sigma_0$ shown in Fig.~\ref{fig:xsec} reflects 
the uncertainty on the OS function $\delta f$ (see the analysis of Ref.~\cite{KP04}).

We now discuss the normalization of the isovector valence quark distribution.
The cancellation of nuclear effects for this quantity is driven by the conservation of the 
vector current (CVC) and the corresponding sum rule reads
\begin{equation}\label{n1:sr}
N_1^A = A^{-1} \int_0^A\ud x\,q_{1/A}^- = \beta
\end{equation}
where $q_1^-=u^- -d^-$. This sum rule becomes trivial
for an isoscalar nucleus with $N_1^A=0$.

Similarly to the isoscalar case discussed above,
we find that the corrections owing to nuclear binding and Fermi motion cancel out in \eq{n1:sr},
while the corrections owing to both the OS effect and coherent multiple scattering, remain finite,
\begin{eqnarray}
\label{N1:os}
\delta N_{1}^{\mathrm{OS}} &=& \beta \average{v}_1\int_0^1 \ud x\, q_{1/p}^-(x) \delta f(x), 
\\
\label{N1:coh}
\delta N_{1}^{\text{coh}} &=&  \int_0^1\ud x\,q_{1/p}^-(x)\delta \mathcal{R}_1^- ,
\end{eqnarray}
where $\average{v}_1=\average{p^2{-}M^2}_1/M^2$ is the nucleon virtuality averaged with
the isovector nuclear spectral function and $\delta\mathcal R_1^-$ is the coherent
nuclear correction to the isovector valence quark distribution given by \eq{coh:1mn}.

To fulfill the normalization condition given by \eq{n1:sr} for a nonisoscalar nucleus 
with $\beta\not=0$, we require an exact cancellation between the OS and the coherent nuclear correction:
\begin{equation}
\label{nbalance1}
N_{1}^\text{OS} + N_{1}^\text{coh}=0.
\end{equation}
Similarly to the isoscalar case in \eq{nbalance}, we use \eq{nbalance1} to constrain the unknown amplitude
$a_1^-$ in the isovector channel. Using \eq{coh:1mn} we observe that the isovector coherent correction
$\delta\mathcal{R}_1^-$ is driven by the isoscalar cross section $\sigma_0^+$ and by the interference of the phases
$\alpha_0^+$ and $\alpha_1^-$.
We use the isovector spectral function of Ref.\cite{KP04} to calculate $N_{1}^\text{OS}$
and then obtain $\sigma_0^+(Q^2,\alpha_1^-)$ by solving \eq{nbalance1}. Then we verify that both solutions, the solution
to \eq{nbalance} and that to \eq{nbalance1}, agree withing $1\sigma$ error band for all values of $Q^2>4\ \gevsq$
at $\alpha_1^-=1.73$. 

To constrain the amplitude $a_1^+$ (the parameter $\alpha_1^+$) we use an equation similar to
\eq{nbalance1}, with the $q_1^-$ distribution replaced with $q_1^+$ and follow a similar procedure.
We find $\alpha_1^+=1.46$.

\subsection{Light-cone momentum sum rule}
\label{sec:msr}
 
The energy-momentum conservation causes the light-cone momentum sum rule at two different levels.
At the hadronic level,
the nuclear light-cone momentum is shared between nucleons and mesons and we have \eq{balance:eq}.
At the partonic level, the light-cone momentum is balanced among quarks, antiquarks, and gluons,
\begin{equation}
\label{eq:msr}
x_{q/A} + x_{\bar q/A} + x_{g/A} = M_A/AM,
\end{equation}
where a sum over different quark flavors is assumed and $x_{a/A}=\int_0^{M_A/A} \ud x\, x q_{a/A}(x,Q^2)/A$
for the quark distribution of flavor $a$. Similar equations hold for antiquarks and gluons.
We recall that the Bjorken variable is defined as $x=Q^2/(2Mq_0)$,
where $M$ is the mass of isoscalar nucleon and $q_0$ is taken in the target rest frame.
For the proton (neutron) target the right hand side of \eq{eq:msr} is trivially equal to unity
(neglecting a small difference in the proton and neutron masses).
Note that \eq{eq:msr} involves the $C$-even and isoscalar combination of quark distributions.
Using the notation $x_a^+ = x_a+x_{\bar a}$ we have
\begin{equation}
\label{eq:xApl}
x_{a/A}^+ = \average{y}_N x_{a/N}^+ +
	    \delta^\text{OS}x_{a}^+ +\delta^\text{mes}x_{a}^+ +\delta^\text{coh}x_{a}^+,
\end{equation}
where the first term on the right is the IA contribution with $\average{y}_N$,
the nucleon fraction of the nuclear light-cone momentum by \eq{y:ia}, and $x_{a/N}^+$
the corresponding momentum of the nucleon. The correction terms are attributed to OS,
MEC, and nuclear shadowing effects which read as 
\begin{subequations}\label{eq:deltax}
\begin{align}
\delta^\text{OS} x_{a}^+ &= \average{y}_N \average{v}_0 \int_0^1\ud x\,xq_{a/N}^+(x,Q^2)\delta f(x), 
\\
\delta^\text{mes}x_{a}^+ &= \average{y}_M x_{a/M}^+ ,
\\
\delta^\text{coh}x_{a}^+ &= \int_0^1 \ud x\,xq_{a/N}^+(x,Q^2)\delta\mathcal R_0^+,
\end{align}
\end{subequations}
where $\average{y}_M$ is the meson fraction of the nuclear light-cone momentum by \eq{y:M}
and $x_{a/M}^+$ is mean momentum of $C$-even quark distribution in mesons. 
We summarize these corrections for several different nuclei 
in Table~\ref{tab:x}. In particular, in this table we list the relative values $\delta x^+/x^+_N$ for each of 
the terms in \eqs{eq:deltax}, together with the total $q+\bar q$ momentum of the proton, $x^+_N$. 
The results of Table~\ref{tab:x} were obtained assuming that the relative shadowing correction 
$\delta\mathcal R^+$ is similar for light and heavy quarks.
We observe a partial cancellation between different nuclear corrections in the total quark  momentum $x_{q/A}^+$.
The resulting nuclear correction to the average $x^+$ turns out to be significantly smaller than
the amplitude of the corresponding correction to the quark distributions in different regions of $x$
(see Sec.\ref{sec:qqbar} for more details).

The sum rule (\ref{eq:msr}) allows us to evaluate the average nuclear gluon momentum
as $x_{g/A}=M_A/AM-x_{q/A}^+$.
As shown in Table~\ref{tab:x}, our results  
indicate an enhancement of gluons in heavy nuclei.
In terms of the average $x$ the gluon enhancement is about $1.5-2$\% at $Q^2=20\,\gevsq$.
The enhancement of nuclear gluon momentum also suggests a gluon antishadowing in nuclei at large values of $x$,
 to compensate the nuclear gluon shadowing effect at small $x$ \cite{Frankfurt:1990xz}.

We also found that the ratio $x_{q/A}^+/x_{q/N}^+$ gradually increases with $Q^2$. 
This behavior is explained by a decreasing fraction of the (negative) nuclear shadowing correction in the numerator.
As a result, according to \eq{eq:msr}, the gluon ratio $x_{g/A}/x_{g/N}$ decreases with $Q^2$.
This effect may indicate that the effect of nuclear shadowing for gluons is increasing with $Q^2$. 
A more detailed discussion of nuclear effects on the gluon distribution goes beyond the scope of the present paper
and will be addressed elsewhere.

\begin{table}[ht]
\begin{center}
\begin{tabular}{l||l|l|l|l|l|l}
Nucleus \   &\ $\average{y}_N$ \ 
            &\ ${\delta^\text{OS} x^+}/{x_N^+}$ \ 
            &\ ${\delta^\text{mes} x^+}/{x_N^+}$ \ 
            &\ ${\delta^\text{coh} x^+}/{x_N^+}$ \  
            &\ $x_{q/A}^+/x_{q/N}^+$  \ 
            &\ $x_{g/A}/x_{g/N}$ \ \\ 
\hline\hline
${}^2$H     &\ 0.9943 \ &\ 0.0058 \ &\ 0.0030 \ &\ -0.0027 \ &\ 1.0004 \ & \ 0.9974 \\
\hline
${}^{12}$C  &\ 0.9718 \ &\ 0.0205 \ &\ 0.0127 \ &\ -0.0217 \ &\ 0.9833 \ & \ 1.0005 \\
\hline
${}^{56}$Fe &\ 0.9656 \ &\ 0.0224 \ &\ 0.0148 \ &\ -0.0336 \ &\ 0.9691 \ & \ 1.0104 \\
\hline
${}^{119}$Sn&\ 0.9638 \ &\ 0.0237 \ &\ 0.0156 \ &\ -0.0403 \ &\ 0.9628 \ & \ 1.0164 \\
\hline
${}^{184}$W &\ 0.9626 \ &\ 0.0250 \ &\ 0.0163 \ &\ -0.0442 \ &\ 0.9596 \ & \ 1.0197 \\
\end{tabular}
\end{center}
\caption{\label{tab:x}
Different contributions to nuclear light-cone momentum sum rule calculated for
a few different nuclear targets using the PDF set of Refs.~\cite{Alekhin:2006zm,Alekhin:2007fh}
at $Q^2=20\,\gevsq$.
The last two columns show the nuclear $q+\bar q$ and the gluon $x$ relative to corresponding nucleon quantities.
}
\end{table}

\subsection{Nuclear Quark and Antiquark Distributions}
\label{sec:qqbar}
 
\begin{sidewaysfigure}[p]
\centering
\vspace{-2em}%
\includegraphics[width=0.52\textwidth,keepaspectratio=true]{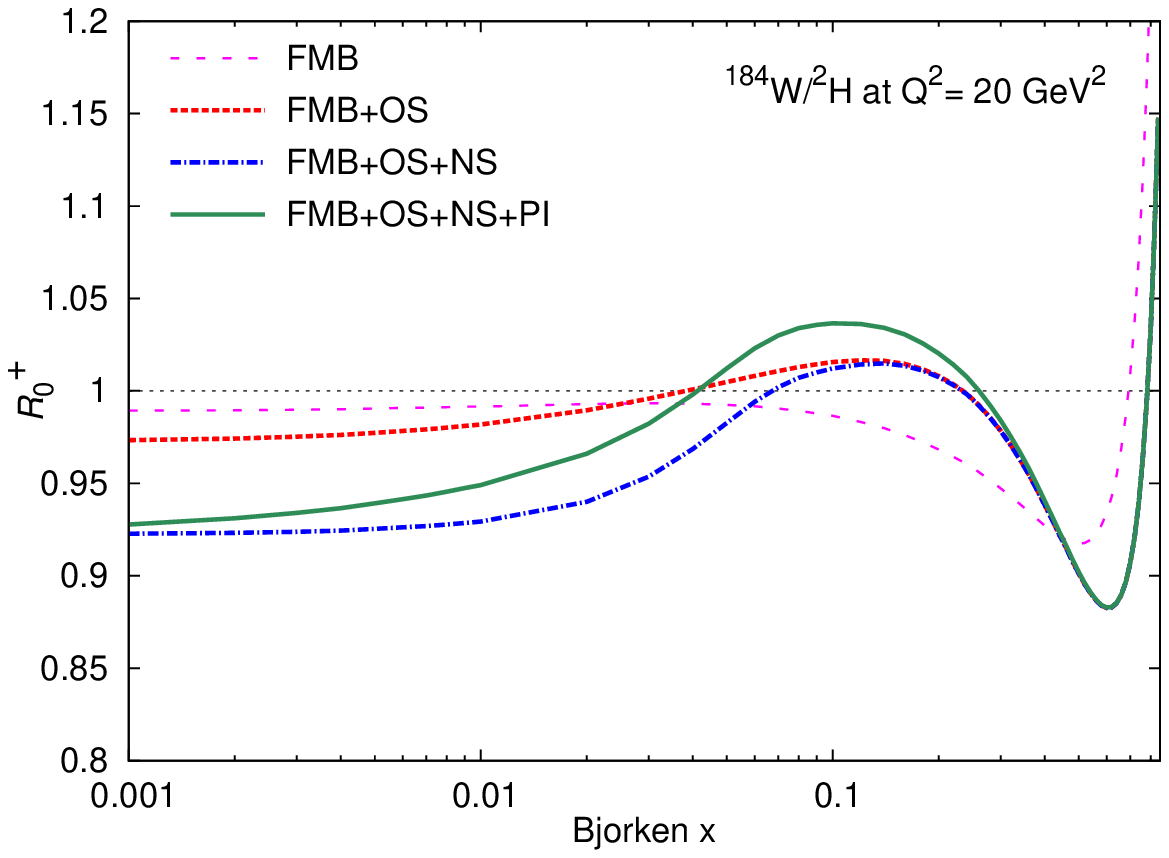}
\includegraphics[width=0.52\textwidth,keepaspectratio=true]{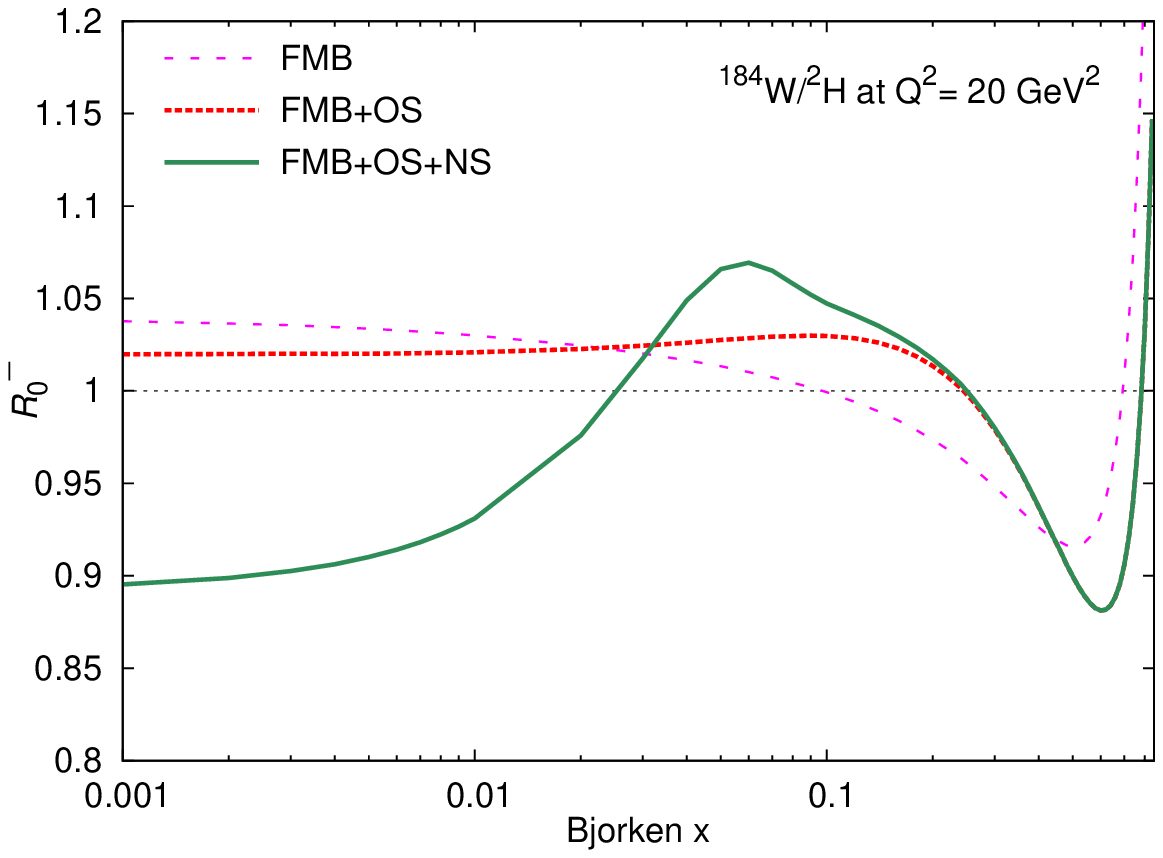}
\includegraphics[width=0.52\textwidth,keepaspectratio=true]{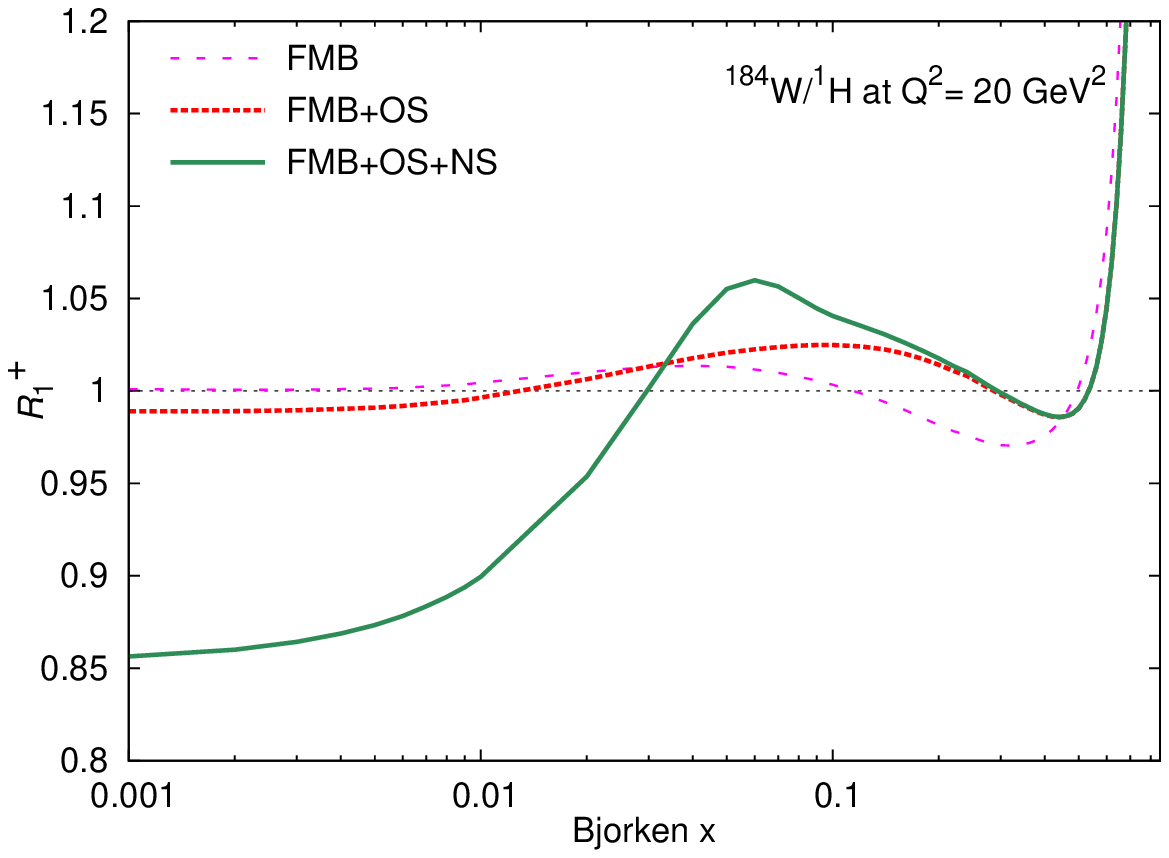}
\includegraphics[width=0.52\textwidth,keepaspectratio=true]{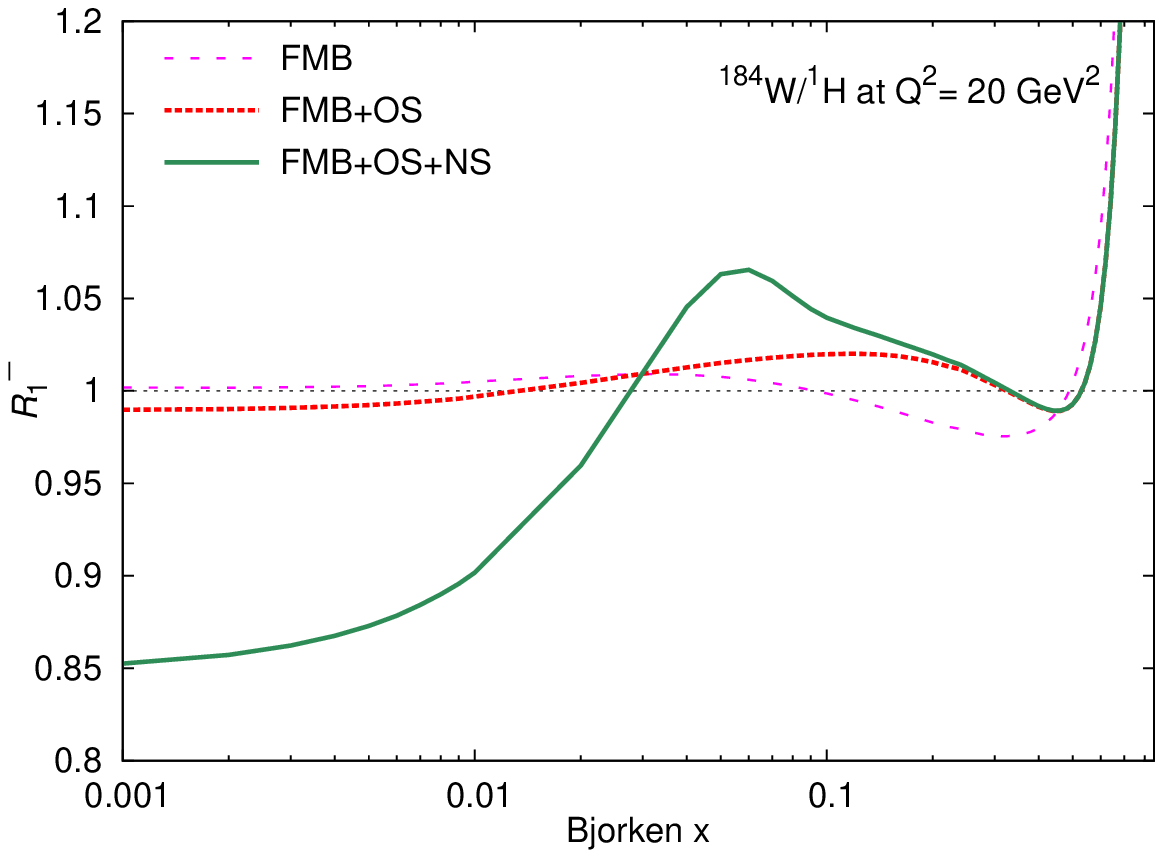}
\caption{\label{fig:npdf}
(Color online)
Different nuclear effects on the $C$-even and $C$-odd combinations of the isoscalar $q_0=u+d$ (top panels)
and the isovector $q_1=u-d$ (bottom panels) quark distributions calculated at $Q^2=20\ \gevsq$.
The ratios are between ${}^{184}$W and the deuteron ${}^2$H (top panels)
and between ${}^{184}$W and the proton ${}^1$H (bottom panel) (see text for details).
}
\end{sidewaysfigure}

In Fig.\ref{fig:npdf} we show different nuclear effects for the $C$-even and the $C$-odd
isoscalar and isovector combinations of the 
quark distributions calculated for the ratio of $^{184}$W and $^2$H nuclei.%
\footnote{%
Note that the light quark contributions to the isoscalar neutrino structure functions $F_2^{\nu+\bar\nu}$
and $F_3^{\nu+\bar\nu}$ are driven by $q_0^+$ and $q_0^-$, respectively.
The isovector combinations (asymmetries) $F_2^{\nu-\bar\nu}$ and $F_3^{\nu-\bar\nu}$ are determined by
$q_1^-$ and $q_1^+$, respectively.
}
The smearing with the nuclear spectral function (Fermi motion and nuclear binding, or FMB),
the OS correction,
the nuclear coherent correction (NS),
and the nuclear meson (PI) correction are treated as discussed in Sec.\ref{sec:npdf}.
In the calculation of FMB we use a nuclear spectral function which takes into account
the mean-field contribution as well as short-range nuclear correlations \cite{KP04}.
Note that the isoscalar and the isovector nuclear spectral functions differ significantly
in Ref.\cite{KP04}.
The OS correction is driven by the function $\delta f(x)$ in \eq{deltaf:def}.
Note that by definition $\delta f$ describes the relative OS effect on a quark distribution
in an off-shell nucleon.
We use the results of Ref.\cite{KP04} and assume a universal OS function $\delta f(x)$,
i.e. same function $\delta f(x)$ for the proton and neutron and for all quark and antiquark
distributions.

\begin{sidewaysfigure}[p]
\centering
\vspace{-2em}%
\includegraphics[width=0.52\textwidth,keepaspectratio=true]{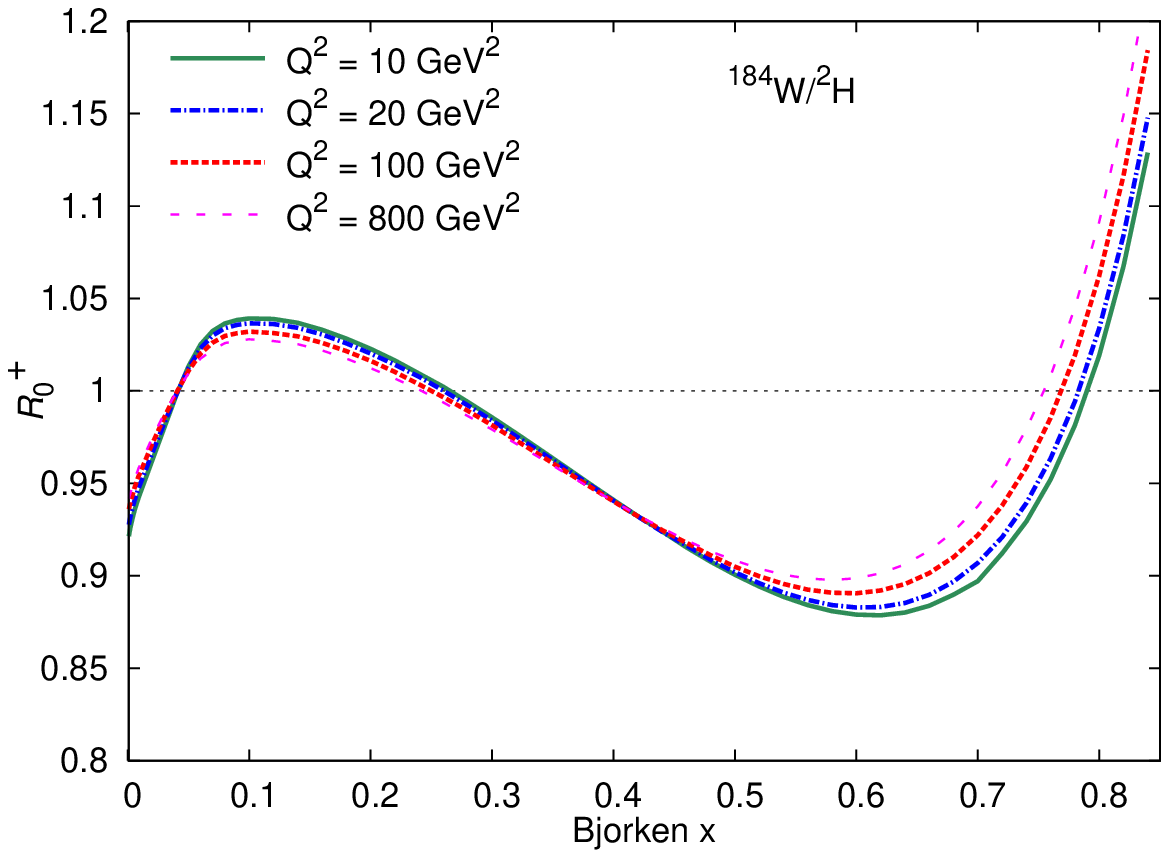}%
\includegraphics[width=0.52\textwidth,keepaspectratio=true]{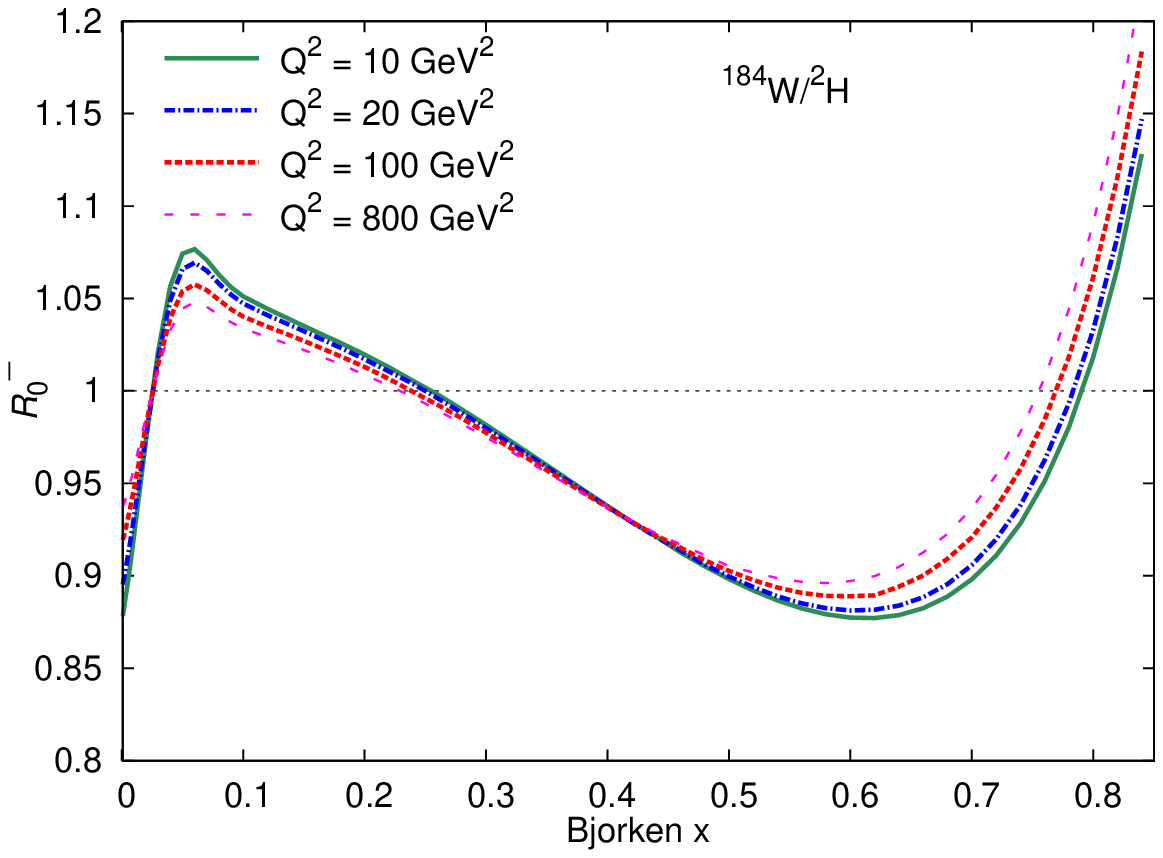}
\includegraphics[width=0.52\textwidth,keepaspectratio=true]{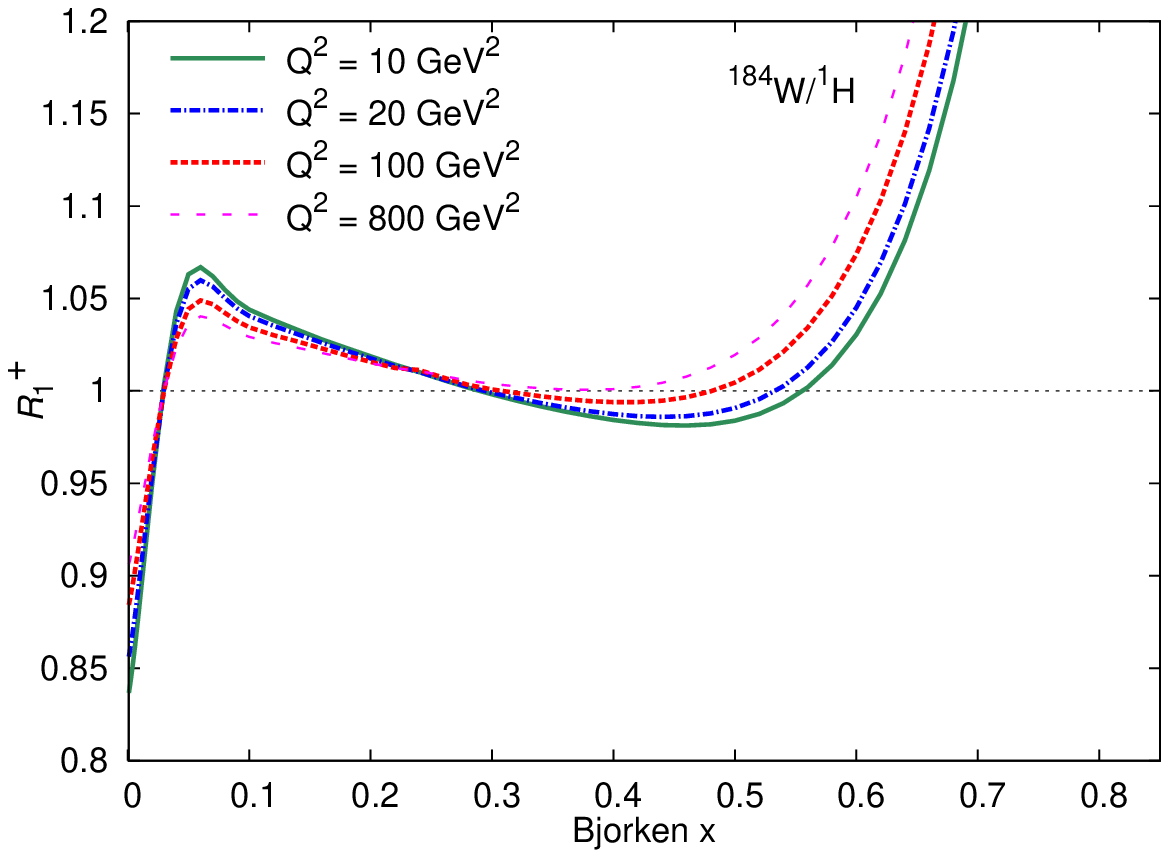}%
\includegraphics[width=0.52\textwidth,keepaspectratio=true]{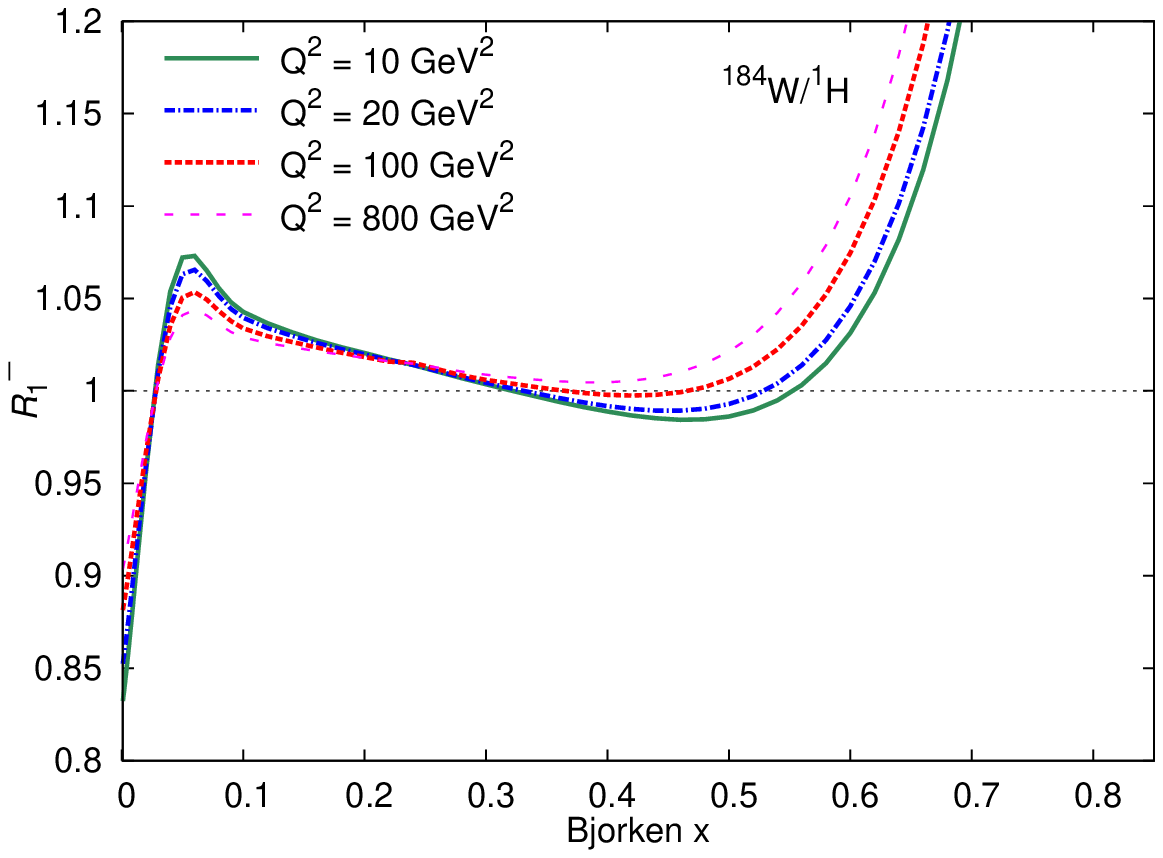}
\caption{\label{fig:npdf-qsq}
(Color online)
The $C$-even and $C$-odd combinations of the isoscalar $q_0=u+d$ (top panels)
and the isovector $q_1=u-d$ (bottom panels) quark distributions
calculated at a few different $Q^2$.
Notations are similar to those in Fig.~\ref{fig:npdf}.
}
\end{sidewaysfigure}

From the top panels of Fig.\ref{fig:npdf} we see that the FMB correction at small $x$
has a different sign for the $C$-even and $C$-odd isoscalar distributions.
This effect is attributable to a significantly different $x$ dependence of $q_0^+$ and $q_0^-$ at low $x$.

At $x<0.01$ the NS correction for the valence quark distribution $q_0^-$ is
enhanced relative to that for $q_0^+$. The underlying reason for
this effect is the enhancement of the multiple scattering corrections for the
cross section asymmetry as discussed in Sec.\ref{npdf:coh}. If
we keep only the double scattering correction, then the ratio $\delta
\mathcal{R}_0^-/\delta \mathcal{R}_0^+$ is given by \eq{r:mn:pl}.
Nevertheless, because of a partial cancellation between FMB and NS corrections for $q_0^-$,
the magnitude of the overall relative nuclear correction at $x<0.01$ 
is similar for valence and sea quarks, being somewhat larger for the former.

Both distributions, the $C$-odd valence $q_0^-$ and the $C$-even $q_0^+$, 
are subject to the antishadowing correction at $x\sim 0.1$.
However, the mechanisms responsible for the antishadowing are different for $q_0^+$ and $q_0^-$.
The enhancement in $q_0^+$ is attributable to the combined effect of the OS and PI corrections.
Instead, the PI correction to $q_0^-$ cancels out, as discussed in Sec.\ref{nuclear:pi}. 
The enhancement in the ratio $\mathcal R_0^-$ is thus entirely attributable to a constructive interference in the
multiple scattering effect from $\Re a_0^-$.

We note that different  nuclear corrections  
on the antiquark distribution $\bar q_0=(q_0^+-q_0^-)/2$
largely cancel out in the antishadowing region. 
In this context we remark that the contribution of the second term in \eq{coh:udbar} becomes increasingly
important at $x>0.05$, because of the ratio $q_0^-(x)/\bar q_0(x)$. This term is negative in that  
region and it partially cancels a positive nuclear pion contribution. 
As a result, the overall nuclear correction to the antiquark distribution is small for
$0.02<x<0.2$. We discuss some implications of this effect in the context of the DY reaction
in Sec.~\ref{sec:nDY}.

At large $x>0.2$ the nuclear corrections to $q_0^+$ and $q_0^-$ are very similar, 
as both distributions are dominated by the valence quarks. 
It should be noted that our result for the relative nuclear correction to  
the valence quark distribution is stable against the specific PDF set chosen 
in the entire region of $x$.
Nuclear effects for sea quarks also depend weakly on the particular choice of PDFs at small values of  
$x$. However, at high $x$ the calculation of nuclear effects for 
antiquark distributions has larger uncertainties and the result is 
sensitive to both the shape and the magnitude of the nucleon antiquark 
distribution.

Nuclear corrections for the isovector quark distribution $q_1=u-d$ are shown in the bottom panels of Fig.~\ref{fig:npdf}
in the form of the ratio $\mathcal R_1=\beta^{-1} q_{1/A}/q_{1/p}$,
where $q_{1/A}$ is the nuclear distribution per nucleon, $q_{1/p}$
is the corresponding distribution in the proton, and $\beta=(Z-N)/A$
is fractional proton excess in a nucleus.
As discussed in Sec.~\ref{sec:npdf}, the isovector nuclear distribution $q_{1/A}$ is
proportional to $\beta$, so that $\beta$ cancels out in the ratio $\mathcal R_1$.
We observe from Fig.~\ref{fig:npdf} that the relative nuclear corrections for the isovector distributions
$q_1^+$ and $q_1^-$ are similar.
Furthermore, the shape and the magnitude of nuclear effects at $x<0.1$ are similar for $q_0^-$ and $q_1^\pm$, and 
they are driven by the coherent nuclear correction discussed in Sec.\ref{npdf:coh}.
At large $x$ the correction is dominated by the nuclear spectral function and by the OS effect. The resulting effect
for the isoscalar channel differs significantly from  that of the isovector channel,
because of the difference between the isoscalar and the isovector nuclear spectral functions \cite{KP04}.
Note that the integral nuclear corrections for the valence distributions $q_{0}^-$ and $q_{1}^-$ are constrained
by the normalization conditions \Eqs{qval:sr}{n1:sr}.

In Fig.\ref{fig:npdf-qsq} we present the results on the same nuclear ratios calculated for different fixed $Q^2$ values.
We observe a weak $Q^2$ dependence of nuclear effects in the $C$-even isoscalar $q_0^+$,
while the $Q^2$ dependence of other distributions is somewhat stronger.

\section{Application to the Drell-Yan process} 
\label{sec:nDY}

The production of lepton pairs with a large mass $Q\gg1\ \gev$ in hadron collisions occurs via the DY process of 
quark-antiquark annihilation (see, e.g., \cite{McGaughey:1999mq,Peng:2014hta}).
The corresponding cross section  depends on the product of the quark and
antiquark distributions in the beam and the target
\begin{equation}\label{DYY}
\frac{\ud^2\sigma^\text{DY}}{\ud x_B \ud x_T} = K \sum_f e_f^2
\left[
q_{f/B}(x_B,Q^2) \bar q_{f/T}(x_T,Q^2) + \bar q_{f/B}(x_B,Q^2) q_{f/T}(x_T,Q^2)
\right],
\end{equation}
where $q_{f/B}$ and $q_{f/T}$ are the quark distributions in the beam and in the target 
and $e_f$ are the quark charges, respectively.
The sum is taken over different quark flavors $f$ and $\bar{q}$ denotes the corresponding antiquark distribution. 
The variables measured experimentally are the mass of the lepton pair $Q$ and the transverse and longitudinal 
momenta of the pair, $k_T$ and $k_L$ respectively. 
The Bjorken variables for the beam and the target, $x_B$ and $x_T$, are related to these quantities 
as $s x_B x_T=Q^2+k_T^2$, with $s$ the total center-of-mass energy squared.
The Feynman variable $x_F=x_B-x_T=2k_L/\sqrt{s}$ depends on the longitudinal momentum of the lepton pair 
in the center-of-mass system.
The factor $K$ in \eq{DYY} absorbs kinematical factors as well as dynamical factors such as 
higher-order QCD corrections.
In this paper we are focused on the analysis of ratios of the DY cross sections for different 
nuclear targets. For this reason we do not write explicitly in \eq{DYY} the factors common to all targets, 
which are canceling out in such ratios.

The proton-induced DY process allows a probe of antiquark distributions in the target and is complementary to the 
lepton-induced DIS. Indeed, in the kinematical region of large $x_B$ and small $x_T$ (large $x_F$) the first 
term in \eq{DYY} dominates and the ratio of the DY yields in different targets is given by the ratio 
of the corresponding antiquark distributions. 
The E772 experiment at Fermilab measured ratios of DY yields originated from the collision of a $800$-$\gev/c$ 
proton beam with five different nuclear targets: ${}^2$H, ${}^{12}$C, ${}^{40}$Ca, ${}^{56}$Fe, 
and ${}^{184}$W \cite{E772}. The DY continuum was studied in the kinematic range $4<Q<9\ \gev$ and $Q>11\ \gev$, 
excluding the quarkonium region, while the Bjorken variable for the target was in the interval $0.04<x_T<0.27$. 
The nuclear dependence of the DY process was also measured by the E866 experiment at Fermilab, 
using the targets ${}^9$Be, ${}^{56}$Fe, and ${}^{184}$W in a similar kinematic region~\cite{E866}.

In addition, the E605 experiment~\cite{Moreno:1990sf} at Fermilab measured the continuum dimuon production by 800~GeV 
protons incident on a copper target in the kinematic range $7 < Q < 9\ \gev$, 
$Q>11\ \gev$, and $0.13 < x_{T} < 0.44$. The published data refer to the absolute DY cross section 
in p-Cu collisions and it is commonly used by global PDF fits~\cite{Alekhin:2012ig,Martin:2009iq,Ball:2012cx,Gao:2013xoa}. 

\subsection{Nuclear effects on Drell-Yan cross section} 
\label{sec:DYnuc}

The nuclear dependence of the DY process comes from two different sources:
 (i) the modification of the (anti)quark distributions in the target nucleus,
 and (ii) the initial state interaction of the projectile 
particle (parton) within the nuclear environment of the target.
We discuss briefly both effects in the following.

We first separate the isoscalar $q_0$ and the isovector $q_1$ contributions in the target in \eq{DYY}.
We have
\begin{equation}\label{DY:01}
\sum_{q=u,d} e_q^2 (q_B \bar q_T + \bar q_B q_T) = 
\sum_{i=0,1} (p_i \bar q_{i/T} + \bar p_i q_{i/T}),
\end{equation}
where $p_0=(4u+d)/18$ and $p_1=(4u-d)/18$, with $u$ and $d$ 
the corresponding quark distributions in the projectile. 
Similar equations can be written for $\bar p_0$ and $\bar p_1$ by replacing  
the quark distributions with the antiquark ones.
In what follows we will discuss the contribution from the isoscalar term.
The isovector correction as well as the contributions from $s$ and $c$ quarks will be addressed elsewhere.

\begin{figure}[htp]
\begin{center}
\includegraphics[width=0.85\textwidth]{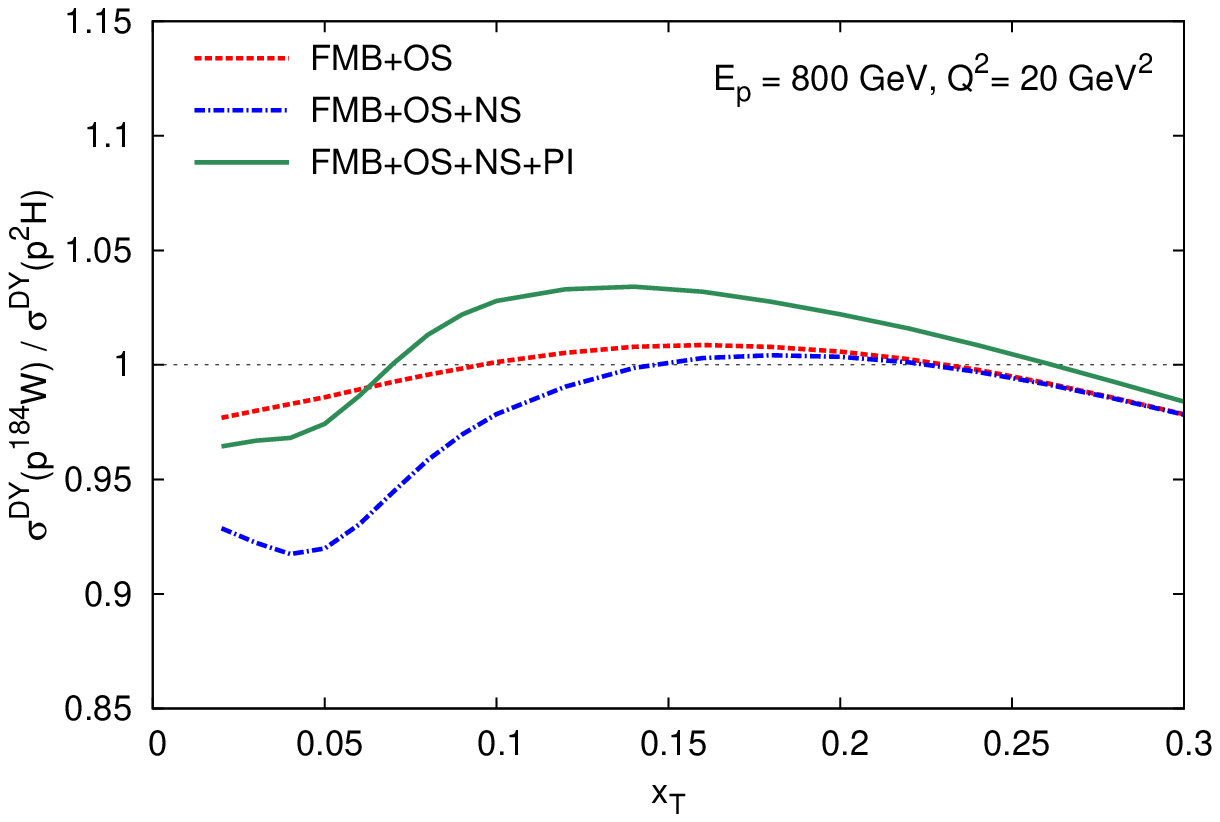}
\includegraphics[width=0.85\textwidth]{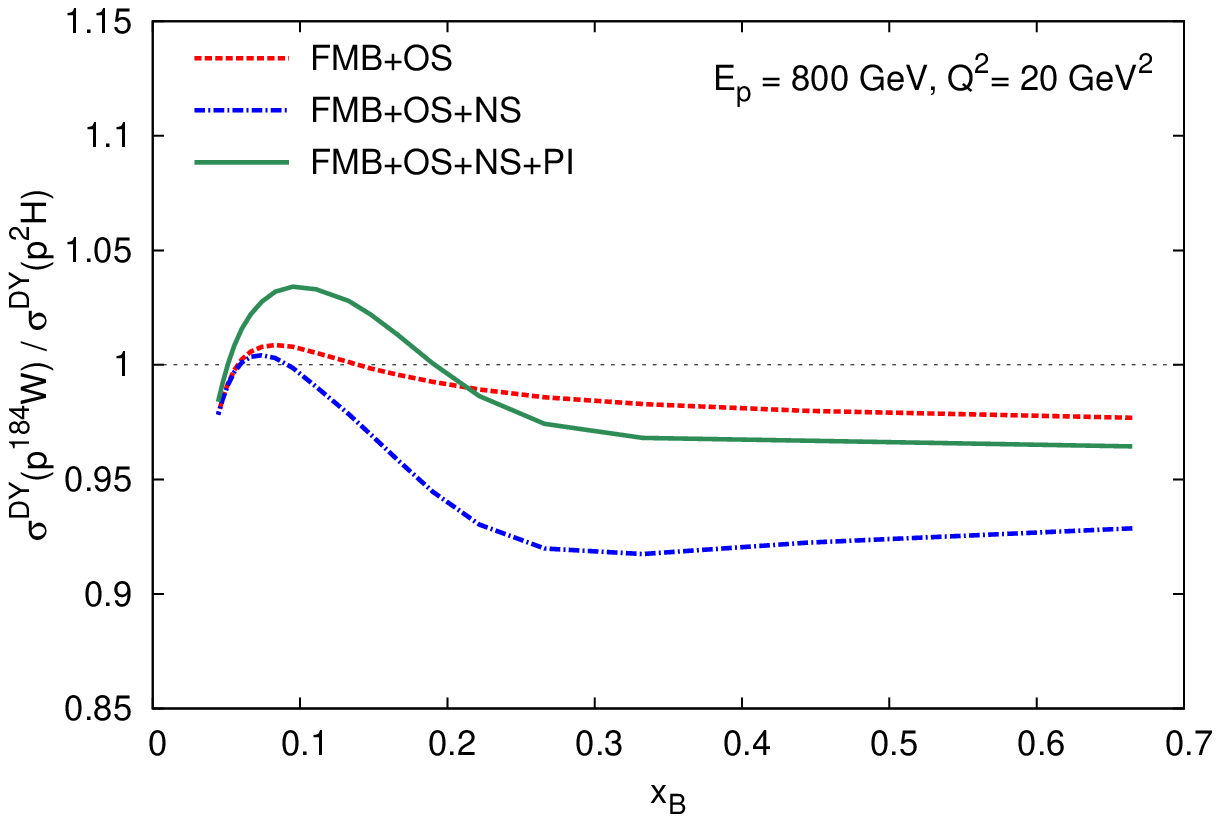}
\caption{%
(Color online)
Different nuclear effects on the cross section for DY lepton pair production 
with a fixed invariant mass $Q^2=20\ \gevsq$ in the collision of $800\ \gev/c$ protons with nuclear targets.
The top panel shows the ratio of the reaction yields in \eq{DYY} between tungsten $^{184}$W and 
deuterium $^2$H targets as a function of $x_T$, while the bottom panel shows the same ratio
as a function of $x_B$.
}
\protect\label{fig:dynuc}
\end{center}
\end{figure}

Nuclear effects on quark and antiquark distributions are discussed in Sec.~\ref{sec:npdf}.
Using those results we calculate the ratio of the DY cross sections on a heavy target and on the deuteron. 
Figure~\ref{fig:dynuc} shows the results obtained at the fixed $Q^2=20\ \gevsq$ and with the variables 
$x_T$ and $x_B$ bound by the relation $s x_T x_B=Q^2$ with $s=1600\ \gevsq$, corresponding 
to the beam energy of the E772 and E866 experiments.
Note that the DY ratios in the region of small $x_T<0.15$ are mainly driven by the corresponding 
ratios of the antiquark distributions. They receive two competing contributions: 
(i) a positive correction due to the nuclear meson exchanged currents (see Sec.\ref{nuclear:pi}) 
and (ii) a negative correction due to nuclear shadowing (see Sec.\ref{npdf:coh}).
These two effects partially cancel out in the antiquark distributions. 
It should be noted that the shadowing correction for antiquarks extends up to a relatively
large $x_T\sim 0.1$. This fact occurs because of the factor $q_{\mathrm val}/\bar q$ in \eq{coh:udbar}, 
which enhances the relative shadowing correction for antiquarks at increasing $x$.
However, such an enhancement is not present for the $q^\pm =q\pm \bar q$ combinations,
 as can be seen from Fig.\ref{fig:npdf}.

The projectile partons in the initial state can undergo multiple soft collisions and can radiate gluons before
annihilating with the (anti)quarks of the target and producing a dimuon pair.
Because of this effect, \eq{DYY} may not be directly applicable to the nuclear DY process,  
as we have to take into account the effects of the propagation of the projectile partons within the 
nuclear environment.
A number of different approaches are available in literature to describe the propagation effects and the 
corresponding gluon radiation in the nuclear medium (for a review see, e.g., Refs.~\cite{Garvey:2002sn,Accardi:2009qv}).
However, results from different analyses significantly disagree both on the magnitude of the quark energy loss
and on its energy and propagation length dependencies \cite{Accardi:2009qv}. 
In this paper we follow the heuristic approach of modifying the variable $x_B$, 
in order to account for the effect of the quark energy loss \cite{Garvey:2002sn}. 
Let $E'=-\ud E/\ud z$ be the parton energy loss in a nucleus per unit length 
($E' \geq0$). If a parton originated with an energy $E_0$ travels over the distance $L$ 
in the nuclear environment before annihilation, then its energy at the moment of the annihilation 
would be $E_1=E_0-E' L$, which will be used to create the dimuon pair.
Therefore, the effect of the energy loss in the nuclear medium requires a correspondingly 
larger value of the initial Bjorken $x_B$.  
In our analysis we assume that \eq{DYY} can be applied to the case in which a nuclear target is present 
with the simple replacement $x_B \to x_B+ E' L/E_B$, where $E_B$ is the energy of the projectile proton.
Below we present the results of our analysis of the combined effects
originated from the nuclear modifications of the target (anti)quark distributions and from the 
energy loss of the beam partons in the nuclear environment. 
To check the sensitivity to the energy loss in the nuclear medium we consider a range of 
possible values commonly used in the literature for this latter $0\leq E' \leq 1.5$~GeV/fm.
We estimate the average propagation length in the nuclear medium of the projectile partons
as $L=3R/4$, which is an average distance traveled by a projectile in a uniform
nuclear density distribution within a sphere of radius $R$.

\subsection{Comparison with Drell-Yan data} 
\label{sec:DYdata} 

\begin{sidewaysfigure}[p]
\centering
\includegraphics[angle=-90,width=\textwidth]{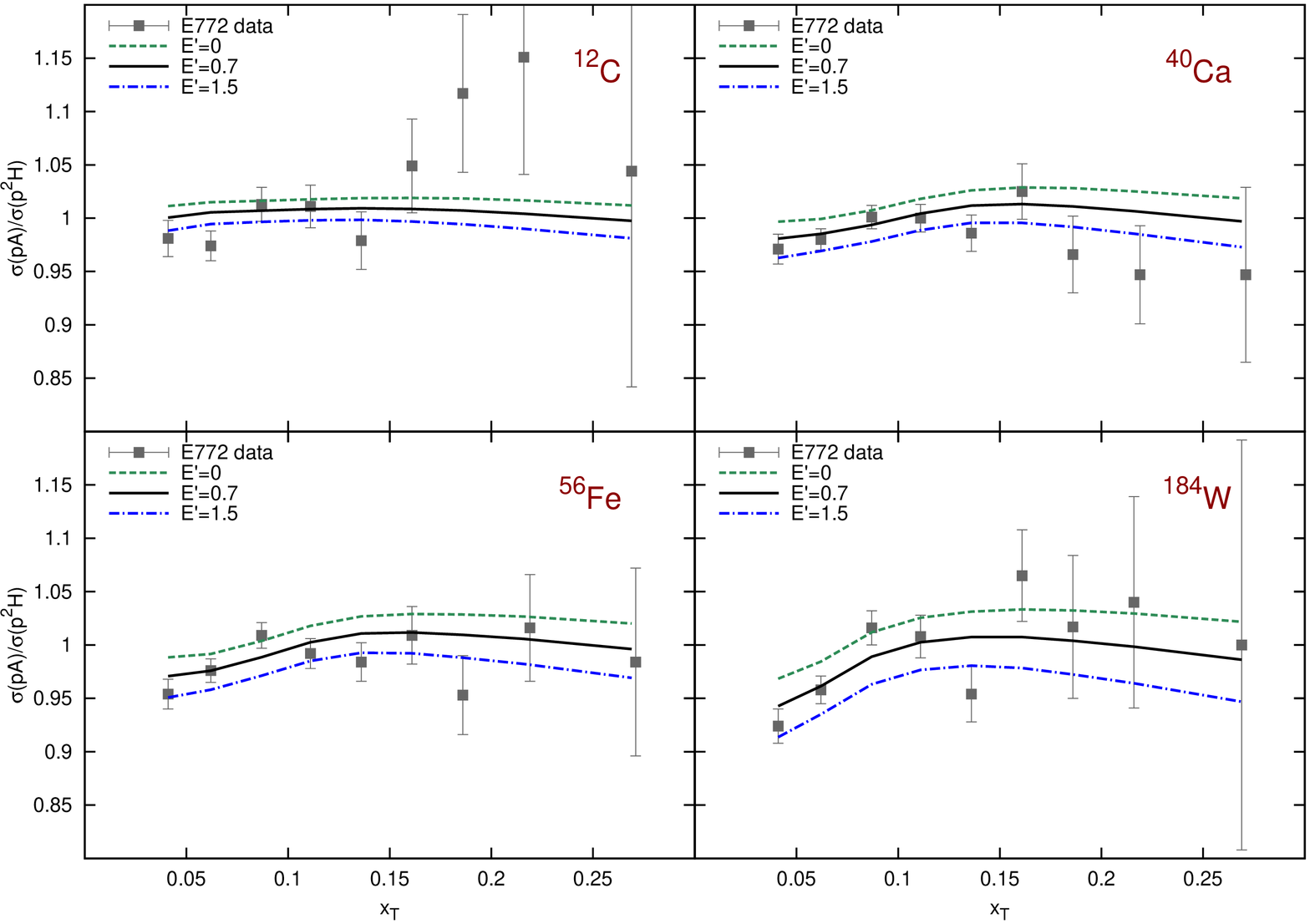}
\caption{%
(Color online)
Ratio of the DY reaction cross sections for different nuclei as a function of $x_T$.
Data points are from the E772 experiment~\cite{E772,dy-web}, while the curves represent our predictions 
with (solid) and without (dashed) the energy loss correction to the projectile quark (see text for details).
\protect\label{fig:e772-xt}}
\end{sidewaysfigure}
\begin{sidewaysfigure}[p]
\centering
\includegraphics[angle=-90,width=\textwidth]{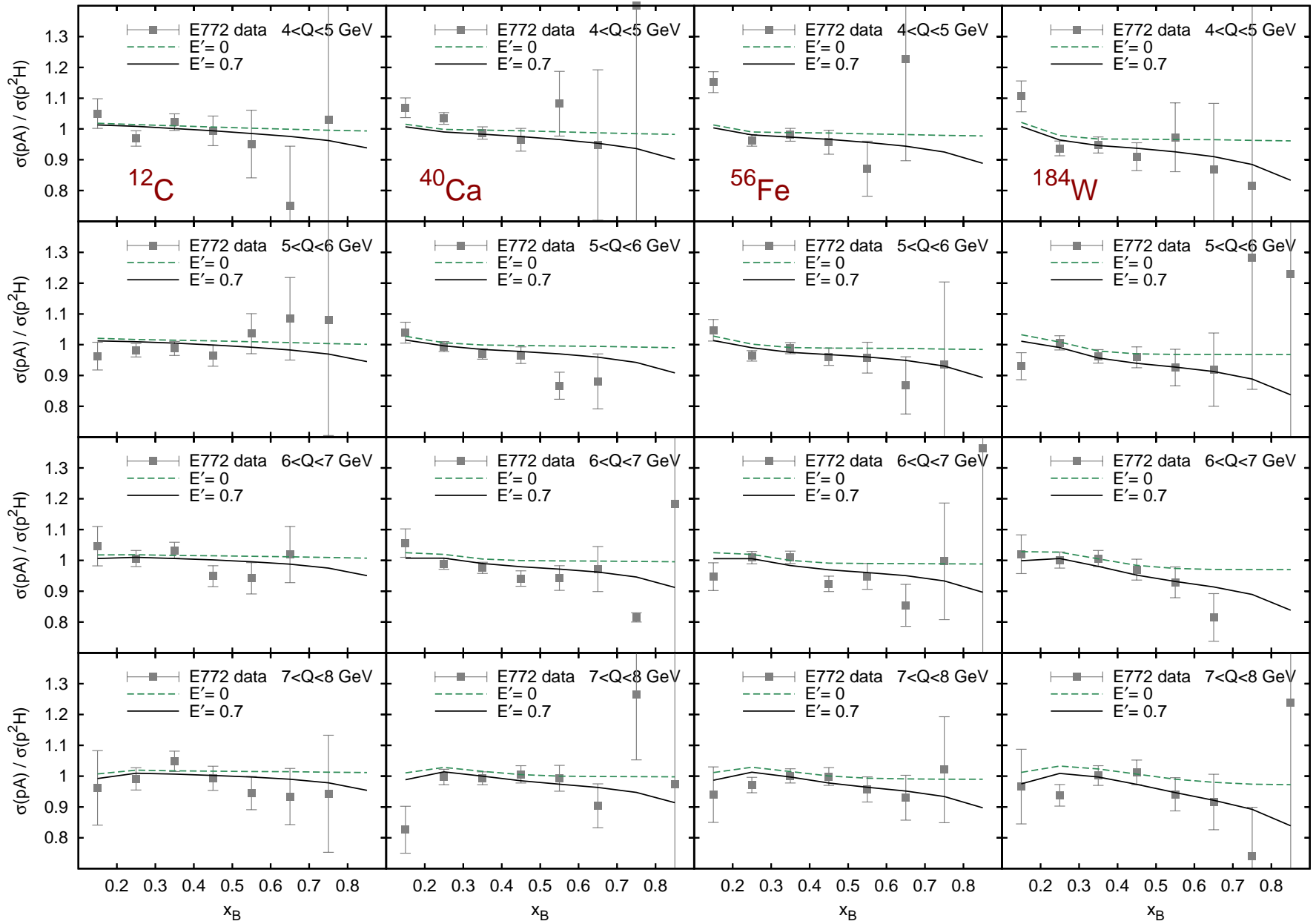}
\caption{%
(Color online)
Ratio of the DY reaction cross sections for different nuclei (rows) and the bins of dimuon mass $Q$ (columns), as a function of $x_B$.
Data points are from the E772 experiment \cite{E772,dy-web}, while the curves represent our predictions 
with (solid) and without (dashed) the energy loss correction to the projectile quark (see text for details).
\protect\label{fig:e772-xb}}
\end{sidewaysfigure}
\begin{sidewaysfigure}[p]
\centering
\includegraphics[angle=-90,width=\textwidth]{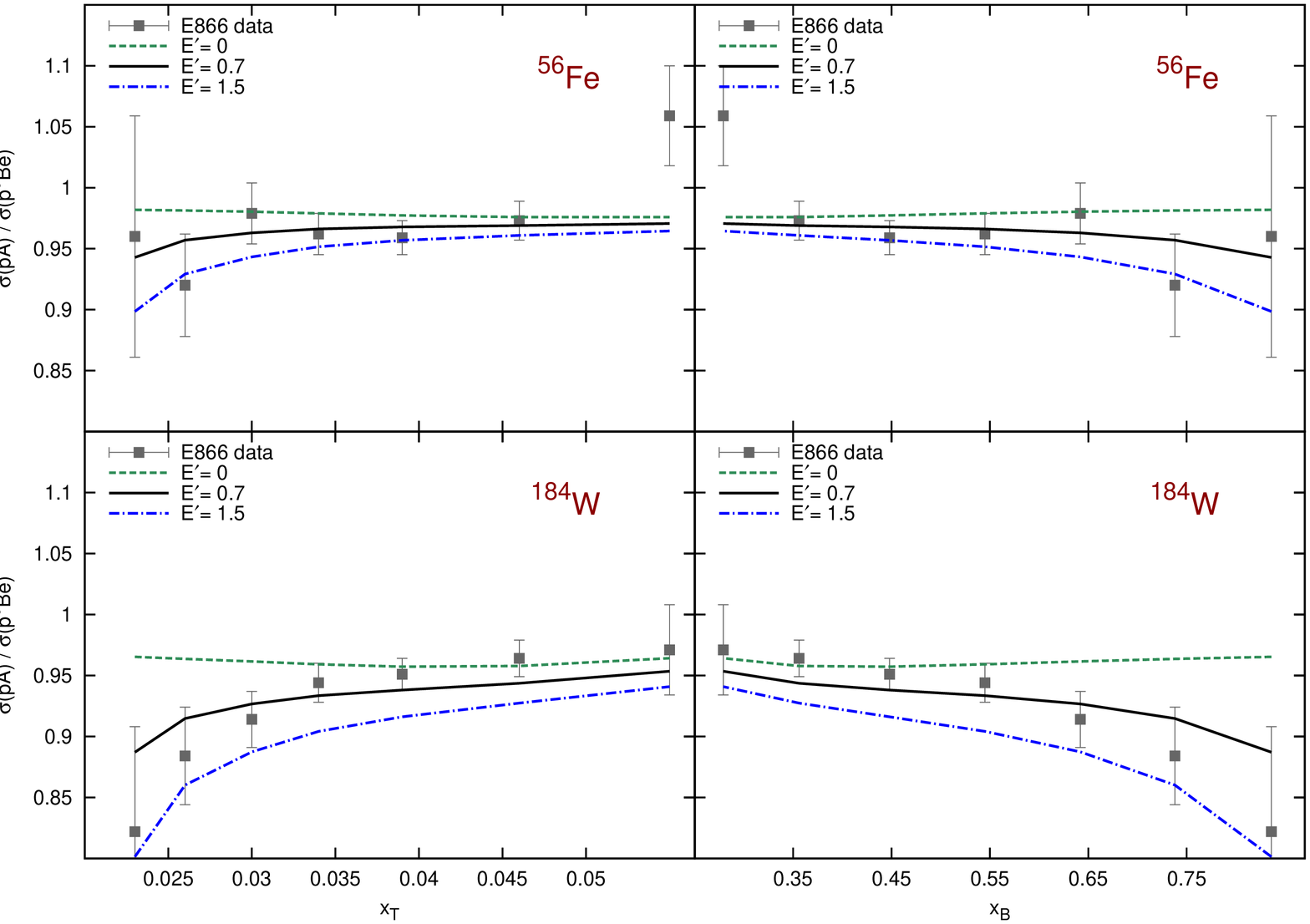}
\caption{%
(Color online)
Comparison of the ratio of the DY reaction yields in different nuclei from the E866 experiment~\cite{E866} with 
our predictions with (solid) and without (dashed) the quark energy loss correction. 
}
\label{fig:e866}
\end{sidewaysfigure}

Figure~\ref{fig:e772-xt} shows a comparison of our predictions with the data from the E772 experiment
for a number of nuclear targets \cite{E772,dy-web}. Although most of the E772 data cover the kinematic 
region in which anti-shadowing is expected according to DIS data ($0.1<x<0.3$), no enhancement is observed 
in the ratio of the DY yields in heavy nuclei and deuterium. This observation gave rise to a long standing 
puzzle because the nuclear binding should result in an excess of nuclear mesons, which is expected to produce 
a marked enhancement in the nuclear anti-quark distributions. However, we found a very good 
agreement of our predictions with the E772 DY data, as illustrated in Fig.~\ref{fig:e772-xt}.
This fact is explained by a partial cancellation between a positive correction owing to the enhancement of the
nuclear meson field and a negative shadowing correction for the antiquark distributions (see Sec.\ref{sec:qqbar}).
Finally, the lowest values of $x_T$ in Fig.\ref{fig:e772-xt} 
clearly show evidence of nuclear shadowing in E772 data, in good agreement with our predictions.

It is worth noting that the good agreement observed with DY data also supports our 
hypothesis of a common OS structure function $\delta f(x)$ for the valence- and the sea-quark distributions.

Figure~\ref{fig:e772-xb} shows the E772 data as a function of $x_B$ for various bins in the 
invariant mass of the dimuon pair \cite{E772,dy-web}, together with our predictions. 
This representation allows a better visualization of the effect 
of the projectile energy loss in the nuclear medium, which is expected to increase with $x_B$.  
The solid curves represent our predictions with a fixed value $E'=0.7$ GeV/fm. The E772 data 
in Figs.\ref{fig:e772-xt} and \ref{fig:e772-xb} favor the presence of moderate energy 
loss effects. Overall, we obtain a very good description of E772 data for both the 
magnitude and the $x$ and mass dependence of the DY cross-section ratios. 
We note that the kinematic coverage of the E772 data is mainly focused on the region 
of intermediate $x_T$ and $x_B$, which is not optimal to address neither the energy 
loss effects nor the nuclear shadowing.    

The data from the E866 experiment~\cite{E866} is shifted towards lower values of $x_T$ and higher 
values of $x_B$ with respect to E772 data, as can be seen from Fig.~\ref{fig:e866}.  
The kinematic coverage of E866 data is therefore focused on the region where both shadowing 
and energy loss effects become more prominent. The E866 data are consistent with the E772 data in the 
overlap region. Figure~\ref{fig:e866} shows that our predictions for the E866 kinematics are in 
good agreement with the E866 data.

We varied the parameter $E'$ describing the parton energy loss 
within the interval from 0 to 1.5~GeV/fm to find its optimal value.
To this end, we evaluated the $\chi^2$ between our predictions and the E772 and E866 data in 
Figs.~\ref{fig:e772-xt} and~\ref{fig:e866}.
The best fit corresponds to a value $E' = 0.70\pm 0.15$ GeV/fm with $\chi^2/d.o.f.=50.8/50$.
The weights of E866 and E772 data in this analysis are comparable because the former 
has a higher sensitivity to energy loss effects, but the latter has more data points available.   
We note that there is a strong correlation in the data between the shadowing correction and the 
energy loss effect, due to the fixed target kinematics, which correlates small values of $x_T$ to 
large values of $x_B$. Furthermore, the kinematic coverage of the available DY data  
is limited to regions in which both effects result in considerable corrections. 

\begin{figure}[htb]
\centering
\includegraphics[width=0.85\textwidth]{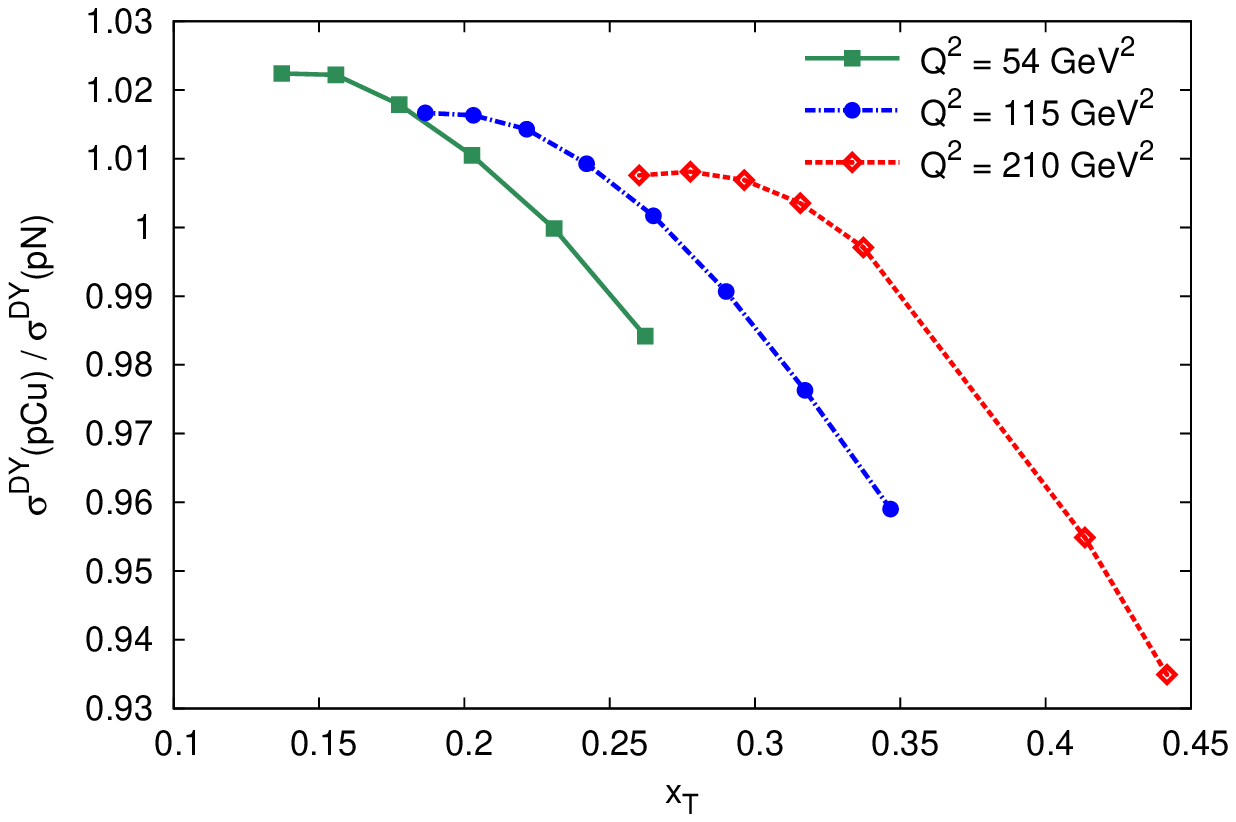}
\caption{%
(Color online)
Ratio of the DY cross sections in the collisions of $800\ \gev$ protons with a ${}^{63}$Cu
target and an average isoscalar nucleon $N=(p+n)/2$, as a function of $x_T$ 
for a few values of the invariant mass square $Q^2$.  
The calculation includes a quark energy loss $E^\prime = 0.7$ GeV/fm. 
\label{fig:E605}}
\end{figure}

We apply the results of our studies of the DY process to calculate the nuclear corrections 
for the dimuon production cross sections measured by the E605 experiment in proton-copper collisions~\cite{Moreno:1990sf}.
Figure~\ref{fig:E605} shows the results for a few fixed invariant masses $Q$ of the dimuon pair.
Table~\ref{tab:E605} lists the nuclear corrections for each E605 data point.
We note that such corrections are relevant for global PDF fits using the data from the  
E605 experiment~\cite{Alekhin:2012ig,Martin:2009iq,Ball:2012cx,Gao:2013xoa}, because 
they can remove the bias introduced by the copper target. 

A new measurement of nuclear effects in the DY production is planned in the experiment 
E906 at Fermilab~\cite{E906}. This experiment will be carried out with a 
120-GeV proton beam and is expected to collect about a factor of 50 larger statistics than that of the  
E772 experiment, using different nuclear targets. The kinematic coverage of E906 data will extend at significantly
larger $x_T$ and should make it possible to disentangle the energy-loss effect from the shadowing corrections.   

\section{Summary} 
\label{sec:discus}

In this article we presented a calculation of nuclear PDFs based on the semi-microscopic model 
of Ref.~\cite{KP04}, focusing at the region of high invariant momentum transfer $Q$.
We discussed in details the $C$-even and $C$-odd combinations of the isoscalar $q_0=u+d$ and 
the isovector $q_1=u-d$ distributions and found a substantial dependence of 
nuclear effects on both the $C$ parity and the isospin
of the PDFs.

In the region at $x>0.2$ nuclear PDFs are dominated by the incoherent contribution
of bound protons and neutrons and the nuclear corrections are driven by the effects of the 
nuclear spectral function together with the OS correction.
The slopes of the EMC effect in different nuclei for $0.3<x<0.6$ 
are explained by the interplay of the nuclear binding, Fermi motion and OS corrections.
We observe a substantial difference in the magnitude of the resulting effect for $q_0$ and $q_1$, 
mainly owing to the difference in the isoscalar and the isovector spectral functions.

All nPDFs show an antishadowing enhancement in the region $0.03<x<0.3$ and a shadowing 
suppression at $x<0.03$.
However, the antishadowing effects for the $q_0^+$ and $q_0^-$ distributions are driven by
different mechanisms. The enhancement in $q_0^+$ is a combined effect of the OS
and of the nuclear MEC corrections, while the antishadowing in $q_0^-$ is attributable to the constructive interference
from the real part of the effective scattering amplitude in the nuclear multiple-scattering series.

The relative correction of the nuclear shadowing is enhanced for the valence-quark distributions ($C$-odd)
and also for the isovector combinations. This effect follows from the corresponding enhancement of the 
propagation effects in the nuclear environment. 
We note that at small $x$ the combined effect of the nuclear binding and of the Fermi motion corrections
has a different sign for the $q_0^+$ and $q_0^-$ distributions, 
owing to the different $x$ dependence of those distributions.
The OS correction is negative at $x<0.02$ for both distributions.
We found a partial cancellation between the nuclear binding and the OS effects in the valence-quark
distribution $q_0^-$ at small $x$. However, both corrections are negative and somewhat enhance the shadowing effect
in the $q_0^+$ distribution.
Overall, the shadowing effect for the $q_0^-$ distribution is more pronounced.
We observe a similar behavior for the isovector quark distributions $q_1^\pm$. 
We also find a weak $Q^2$ dependence of nuclear effects of the $C$-even isoscalar $q_0^+$,
while the $Q^2$ dependence of other distributions is somewhat stronger.

The PDF normalization conditions and the energy-momentum sum rules link nuclear effects
of different origin located in different kinematical regions of $x$. 
In particular, we used the normalization conditions for the isoscalar and 
the isovector valence quark distributions as equations to determine the unknown amplitudes 
controlling the coherent nuclear correction. We then solved these equations in terms of the OS
correction to the corresponding distributions.
We also use the light-cone momentum sum rule together with equations of motion
to constrain the mesonic light-cone distributions and 
calculate the corresponding mesonic corrections to nuclear PDFs.
As a result, in this approach the nuclear modifications to PDFs are essentially determined by the nuclear spectral function, 
and by the OS function of the nucleon $\delta f$.

We applied our model of nPDFs to calculate the cross-sections for DY production 
in proton-nucleus collisions. 
We recall that the E772 data on ratios of DY yields in different nuclear targets
show no enhancement of the nuclear sea quark distributions in the antishadowing region $x \sim 0.05-0.2$.
This behavior is in contrast with the enhancement of the nuclear sea due to the nuclear meson contributions.
We found that this discrepancy can be explained by a partial cancellation between different 
nuclear corrections on the antiquark distributions in the antishadowing region.
Our predictions are in very good agreement with both the 
magnitude and the $x$ and mass dependence of the DY data from the E772 and E866 experiments \cite{E772,E866}.
We also discussed the impact of the energy loss of the projectile partons
in the nuclear environment on the ratio of the DY yields in different nuclear targets
and found that our analysis favors an energy loss around $0.7\ \text{GeV}/\text{fm}$.
We applied the results of our studies to calculate the nuclear corrections 
for the dimuon production cross section data measured by the E605 experiment in proton-copper collisions, 
which are relevant to remove the bias introduced by the nuclear target in global PDF fits.

In the study of nPDFs we assumed that $\delta f$ is universal and flavor-independent \cite{KP04,KP10}. 
The good agreement of the corresponding predictions with the available DY data is 
consistent with a common OS function for valence- and sea-quark distributions. 
We also remark that the available DIS and DY data have a limited sensitivity
to isospin effects and, therefore, cannot address a possible isospin dependence of $\delta f$.

As a final remark we note that the studies of nuclear effects in the isovector combinations
of (anti)quark distributions $q_1$ are important for a calculation of the effects of 
the neutron excess in heavy nuclei for high-energy nuclear reactions, including DIS and DY production.
The isovector distributions are also of direct relevance for neutrino physics,
as they determine the $\nu-\bar\nu$ asymmetries in neutrino-nuclear collisions
\cite{Kulagin:2003wz,Kulagin:2004jn,KP07}. 
In particular, an accurate knowledge of such effects is crucial for the interpretation of 
data in modern neutrino experiments~\cite{Adams:2013qkq,Mishra:2008nx}.
This, in turn, requires detailed studies of the isovector component of the nuclear spectral function
as well as the isospin dependence of the OS function $\delta f$.

\section*{Acknowledgements}

We thank G.A. Miller for interesting discussions on the topics covered in this work.
We would like to thank Center for Theoretical Underground Physics and Related Areas (CETUP*)
for its hospitality and partial support during the 2014 Summer Program.
S.K. was supported by Russian Science Foundation Grant No. 14-22-00161.
R.P. was supported by Grant No. DE-FG02-13ER41922 from the Department of Energy, USA. 

\newpage
\twocolumngrid
\appendix
\section{Nuclear corrections for the E605 experiment}
\label{apndx:E605}
\tablehead{$x_{F}$ & $\sqrt{\tau}$ & $x_{B}$ & $x_{T}$ & $Q^2$ & $\frac{\sigma (pCu)}{\sigma (pN)}$ \\ \hline\hline }
\tablecaption{\label{tab:E605}
The ratio of dimuon production cross section in proton collisions with a ${}^{63}$Cu target and an average isoscalar nucleon 
$N=(p+n)/2$, as a function of $x_F=x_B-x_T,\ \sqrt{\tau}=\sqrt{x_B x_T},\ x_B,\ x_T,\ Q^2$. 
Each line corresponds to a data point with the actual kinematics measured by the E605 experiment. The ratios were calculated with the quark energy loss $E^\prime = 0.7$ GeV/fm.
}

\footnotesize
\begin{supertabular}{c|c|c|c|c|c} 
-0.125 & 0.1831 & 0.1310 & 0.2560 & 50.47 &  0.9859 \\
-0.125 & 0.1897 & 0.1372 & 0.2622 & 54.17 &  0.9842 \\
-0.125 & 0.1974 & 0.1446 & 0.2696 & 58.66 &  0.9822 \\
-0.125 & 0.2038 & 0.1507 & 0.2757 & 62.53 &  0.9806 \\
-0.125 & 0.2117 & 0.1582 & 0.2832 & 67.47 &  0.9785 \\
-0.125 & 0.2188 & 0.1651 & 0.2901 & 72.07 &  0.9761 \\
-0.125 & 0.2264 & 0.1724 & 0.2974 & 77.16 &  0.9739  \\
-0.125 & 0.2338 & 0.1795 & 0.3045 & 82.29  &  0.9717 \\
-0.125 & 0.2772 & 0.2217 & 0.3467 & 115.68 &  0.9590 \\
-0.125 & 0.2847 & 0.2290 & 0.3540 & 122.02 &  0.9568 \\
-0.125 & 0.2917 & 0.2358 & 0.3608 & 128.10 &  0.9548 \\
-0.125 & 0.2994 & 0.2434 & 0.3684 & 134.95 &  0.9526 \\
-0.125 & 0.3069 & 0.2507 & 0.3757 & 141.79 &  0.9505 \\
-0.125 & 0.3201 & 0.2636 & 0.3886 & 154.25 &  0.9470 \\
-0.125 & 0.3428 & 0.2860 & 0.4110 & 176.91 &  0.9415 \\
-0.125 & 0.3741 & 0.3168 & 0.4418 & 210.69 &  0.9349 \\
-0.075 & 0.1831 & 0.1494 & 0.2244 & 50.47 &  0.9998 \\
-0.075 & 0.1897 & 0.1559 & 0.2309 & 54.17 &  0.9998  \\ 
-0.075 & 0.1974 & 0.1634 & 0.2383 & 58.66 &  0.9976 \\
-0.075 & 0.2038 & 0.1697 & 0.2447 & 62.53 &  0.9959 \\
-0.075 & 0.2117 & 0.1775 & 0.2525 & 67.47 &  0.9937 \\
-0.075 & 0.2188 & 0.1845 & 0.2595 & 72.07 &  0.9918 \\
-0.075 & 0.2264 & 0.1920 & 0.2670 & 77.16 &  0.9898 \\
-0.075 & 0.2338 & 0.1993 & 0.2743 & 82.29 &  0.9878 \\
-0.075 & 0.2772 & 0.2422 & 0.3172 & 115.68 &  0.9763 \\
-0.075 & 0.2847 & 0.2497 & 0.3247 & 122.02 &  0.9744 \\
-0.075 & 0.2917 & 0.2566 & 0.3316 & 128.10 &  0.9726 \\
-0.075 & 0.2994 & 0.2642 & 0.3392 & 134.95 &  0.9707 \\
-0.075 & 0.3069 & 0.2717 & 0.3467 & 141.79 &  0.9689 \\
-0.075 & 0.3201 & 0.2848 & 0.3598 & 154.25 &  0.9658 \\
-0.075 & 0.3428 & 0.3073 & 0.3823 & 176.91 &  0.9608 \\
-0.075 & 0.3741 & 0.3385 & 0.4135 & 210.69 &  0.9549 \\
-0.075 & 0.3993 & 0.3636 & 0.4386 & 240.03 &  0.9511 \\
-0.025 & 0.1825 & 0.1704 & 0.1954 & 50.14 &  1.0120  \\
-0.025 & 0.1897 & 0.1776 & 0.2026 & 54.17 &  1.0105  \\
-0.025 & 0.1969 & 0.1848 & 0.2098 & 58.37 &  1.0090  \\
-0.025 & 0.2041 & 0.1920 & 0.2170 & 62.71 &  1.0074  \\
-0.025 & 0.2116 & 0.1995 & 0.2245 & 67.41 &  1.0058  \\
-0.025 & 0.2192 & 0.2071 & 0.2321 & 72.33 &  1.0041  \\
-0.025 & 0.2264 & 0.2142 & 0.2392 & 77.16 &  1.0024  \\
-0.025 & 0.2336 & 0.2214 & 0.2464 & 82.15 &  1.0007  \\
-0.025 & 0.2773 & 0.2651 & 0.2901 & 115.76 & 0.9907 \\
-0.025 & 0.2844 & 0.2722 & 0.2972 & 121.76 &  0.9891 \\
-0.025 & 0.2917 & 0.2795 & 0.3045 & 128.10 &  0.9875 \\
-0.025 & 0.2987 & 0.2865 & 0.3115 & 134.32 &  0.9861 \\
-0.025 & 0.3064 & 0.2942 & 0.3192 & 141.33 &  0.9845 \\
-0.025 & 0.3199 & 0.3076 & 0.3326 & 154.06 &  0.9818 \\
-0.025 & 0.3430 & 0.3307 & 0.3557 & 177.11 &  0.9775 \\
-0.025 & 0.3760 & 0.3637 & 0.3887 & 212.83 &  0.9722 \\
-0.025 & 0.4044 & 0.3921 & 0.4171 & 246.20 &  0.9688 \\
0.025  & 0.1825 & 0.1954 & 0.1704 & 50.14 &  1.0191  \\
0.025  & 0.1897 & 0.2026 & 0.1776 & 54.17 &  1.0179  \\
0.025  & 0.1969 & 0.2098 & 0.1848 & 58.37 &  1.0166  \\
0.025  & 0.2041 & 0.2170 & 0.1920 & 62.71 &  1.0153  \\
0.025  & 0.2116 & 0.2245 & 0.1995 & 67.41 &  1.0140  \\
0.025  & 0.2192 & 0.2321 & 0.2071 & 72.33 &  1.0127  \\
0.025  & 0.2264 & 0.2392 & 0.2142 & 77.16 &  1.0115  \\
0.025  & 0.2336 & 0.2464 & 0.2214 & 82.15 &  1.0102  \\
0.025  & 0.2773 & 0.2901 & 0.2651 & 115.76 & 1.0017  \\
0.025  & 0.2844 & 0.2972 & 0.2722 & 121.76 &  1.0004  \\
0.025  & 0.2917 & 0.3045 & 0.2795 & 128.10 &  0.9991 \\
0.025  & 0.2987 & 0.3115 & 0.2865 & 134.32 &  0.9978 \\
0.025  & 0.3064 & 0.3192 & 0.2942 & 141.33 &  0.9965  \\
0.025  & 0.3199 & 0.3326 & 0.3076 & 154.06 &  0.9942 \\
0.025  & 0.3430 & 0.3557 & 0.3307 & 177.11 &  0.9907 \\
0.025  & 0.3760 & 0.3887 & 0.3637 & 212.83 &  0.9865  \\
0.025  & 0.4044 & 0.4171 & 0.3921 & 246.20 &  0.9837 \\
0.075  & 0.1824 & 0.2237 & 0.1487 & 50.09 &  1.0229   \\
0.075  & 0.1896 & 0.2308 & 0.1558 & 54.12 &  1.0222   \\
0.075  & 0.1970 & 0.2380 & 0.1630 & 58.42 &  1.0209  \\
0.075  & 0.2045 & 0.2454 & 0.1704 & 62.96 &  1.0202  \\
0.075  & 0.2116 & 0.2524 & 0.1774 & 67.41 &  1.0192  \\
0.075  & 0.2188 & 0.2595 & 0.1845 & 72.07 &  1.0182   \\
0.075  & 0.2262 & 0.2668 & 0.1918 & 77.03  & 1.0171  \\
0.075  & 0.2333 & 0.2738 & 0.1988 & 81.94 &  1.0161  \\
0.075  & 0.2770 & 0.3170 & 0.2420 & 115.51 &  1.0092  \\
0.075  & 0.2845 & 0.3245 & 0.2495 & 121.85 &  1.0081   \\
0.075  & 0.2915 & 0.3314 & 0.2564 & 127.92 &   1.0071  \\
0.075  & 0.2992 & 0.3390 & 0.2640 & 134.77 &  1.0060   \\
0.075  & 0.3064 & 0.3462 & 0.2712 & 141.33 &  1.0050  \\
0.075  & 0.3199 & 0.3596 & 0.2846 & 154.06 &  1.0031  \\
0.075  & 0.3433 & 0.3828 & 0.3078 & 177.42 &  1.0001  \\
0.075  & 0.3729 & 0.4123 & 0.3373 & 209.34 &  0.9971 \\
0.075  & 0.4010 & 0.4402 & 0.3652 & 242.08 &   0.9949 \\
0.075  & 0.4367 & 0.4758 & 0.4008 & 287.10 &   0.9933 \\
0.125  & 0.1824 & 0.2553 & 0.1303 & 50.09 &   1.0214  \\
0.125  & 0.1896 & 0.2621 & 0.1371 & 54.12 &  1.0224   \\
0.125  & 0.1970 & 0.2692 & 0.1442 & 58.42 &  1.0228  \\
0.125  & 0.2045 & 0.2763 & 0.1513 & 62.96 &  1.0226  \\
0.125  & 0.2116 & 0.2831 & 0.1581 & 67.41 &  1.0218  \\
0.125  & 0.2188 & 0.2901 & 0.1651 & 72.07 &  1.0208  \\
0.125  & 0.2262 & 0.2972 & 0.1722 & 77.03  & 1.0202  \\
0.125  & 0.2333 & 0.3040 & 0.1790 & 81.94 &  1.0194  \\
0.125  & 0.2770 & 0.3465 & 0.2215 & 115.51 &  1.0143  \\
0.125  & 0.2845 & 0.3538 & 0.2288 & 121.85 &   1.0130  \\
0.125  & 0.2915 & 0.3606 & 0.2356 & 127.92 &  1.0120  \\
0.125  & 0.2992 & 0.3682 & 0.2432 & 134.77 &  1.0111  \\
0.125  & 0.3064 & 0.3752 & 0.2502 & 141.33 &  1.0102  \\
0.125  & 0.3199 & 0.3884 & 0.2634 & 154.06 &  1.0087  \\
0.125  & 0.3433 & 0.4114 & 0.2864 & 177.42 &  1.0062  \\
0.125  & 0.3729 & 0.4406 & 0.3156 & 209.34 &   1.0035  \\
0.125  & 0.4010 & 0.4683 & 0.3433 & 242.08 &  1.0018  \\
0.125  & 0.4367 & 0.5036 & 0.3786 & 287.10 &  1.0007  \\
0.175  & 0.2045 & 0.3099 & 0.1349 & 62.96 &  1.0207  \\
0.175  & 0.2120 & 0.3168 & 0.1418 & 67.66 &   1.0216  \\
0.175  & 0.2189 & 0.3232 & 0.1482 & 72.14 &   1.0218  \\
0.175  & 0.2260 & 0.3298 & 0.1548 & 76.89 &  1.0214  \\
0.175  & 0.2334 & 0.3368 & 0.1618 & 82.01 &  1.0203  \\
0.175  & 0.2771 & 0.3781 & 0.2031 & 115.59 &  1.0163  \\ 
0.175  & 0.2843 & 0.3850 & 0.2100 & 121.68 &  1.0157  \\
0.175  & 0.2915 & 0.3918 & 0.2168 & 127.92 &  1.0151  \\
0.175  & 0.2988 & 0.3988 & 0.2238 & 134.41 &  1.0141  \\ 
0.175  & 0.3064 & 0.4061 & 0.2311 & 141.33 &  1.0131  \\
0.175  & 0.3198 & 0.4191 & 0.2441 & 153.96 &  1.0116  \\ 
0.175  & 0.3450 & 0.4434 & 0.2684 & 179.18 &  1.0093   \\
0.175  & 0.3737 & 0.4713 & 0.2963 & 210.24 &  1.0069  \\
0.175  & 0.4054 & 0.5022 & 0.3272 & 247.42 &  1.0052  \\
0.175  & 0.4347 & 0.5309 & 0.3559 & 284.47 &   1.0045   \\
0.225  & 0.2334 & 0.3716 & 0.1466 & 82.01 &  1.0201  \\
0.225  & 0.2771 & 0.4116 & 0.1866 & 115.59 &  1.0167  \\ 
0.225  & 0.2843 & 0.4182 & 0.1932 & 121.68 &  1.0160  \\
0.225  & 0.2915 & 0.4250 & 0.2000 & 127.92 &  1.0154   \\
0.225  & 0.2988 & 0.4318 & 0.2068 & 134.41 &  1.0148  \\
0.225  & 0.3064 & 0.4389 & 0.2139 & 141.33 &  1.0143  \\
0.225  & 0.3198 & 0.4515 & 0.2265 & 153.96 &  1.0128  \\
0.225  & 0.3450 & 0.4754 & 0.2504 & 179.18 &  1.0103  \\
0.225  & 0.3737 & 0.5028 & 0.2778 & 210.24 &  1.0081   \\
0.225  & 0.4054 & 0.5332 & 0.3082 & 247.42 &  1.0062  \\
0.225  & 0.4347 & 0.5615 & 0.3365 & 284.47 &   1.0054  \\
0.275  & 0.2998 & 0.4673 & 0.1923 & 135.31 &  1.0139  \\
0.275  & 0.3062 & 0.4732 & 0.1982 & 141.15 &   1.0134  \\
0.275  & 0.3205 & 0.4862 & 0.2112 & 154.64 &  1.0123  \\
0.275  & 0.3461 & 0.5099 & 0.2349 & 180.33 &  1.0097  \\
0.275  & 0.3731 & 0.5351 & 0.2601 & 209.56 &  1.0076  \\
0.275  & 0.4027 & 0.5630 & 0.2880 & 244.13 &  1.0055  \\
0.275  & 0.4318 & 0.5907 & 0.3157 & 280.69 &  1.0041 \\
\end{supertabular}


\begin{thebibliography}{99}

\bibitem{Collins:1989gx} 
  J.~C.~Collins, D.~E.~Soper and G.~F.~Sterman,
  Adv.\ Ser.\ Direct.\ High Energy Phys.\  {\bf 5}, 1 (1988).

\bibitem{Alekhin:2012ig} 
  S.~Alekhin, J.~Blumlein and S.~Moch,
  Phys.\ Rev.\ D {\bf 86}, 054009 (2012).

\bibitem{Martin:2009iq} 
  A.~D.~Martin, W.~J.~Stirling, R.~S.~Thorne and G.~Watt,
  Eur.\ Phys.\ J.\ C {\bf 63}, 189 (2009).

\bibitem{Ball:2012cx} 
  R.~D.~Ball, V.~Bertone, S.~Carrazza, {\it et al.}, 
  Nucl.\ Phys.\ B {\bf 867}, 244 (2013).

\bibitem{Gao:2013xoa} 
  J.~Gao, M.~Guzzi, J.~Huston {\it et al.},  {\it et al.},
Phys. Rev. D {\bf 89}, 033009 (2014).

\bibitem{Arneodo:1992wf}
  M.~Arneodo,  
  Phys.\ Rept.\ {\bf 240}, 301-393 (1994).

\bibitem{Norton:2003cb} 
  P.~R.~Norton,
  Rept.\ Prog.\ Phys.\  {\bf 66}, 1253 (2003).

\bibitem{Geesaman:1995yd}
  D.~F.~Geesaman, K.~Saito and A.~Thomas,  
  Ann.\ Rev.\ Nucl.\ Part.\ Sci.\ {\bf 45}, 337-390 (1995).

\bibitem{Bickerstaff:1989ch} 
  R.~P.~Bickerstaff and A.~W.~Thomas,
  J.\ Phys.\ G {\bf 15}, 1523 (1989).

\bibitem{Eskola:2009uj} 
  K.~J.~Eskola, H.~Paukkunen and C.~A.~Salgado,
  JHEP {\bf 0904}, 065 (2009).

\bibitem{Hirai:2007sx} 
  M.~Hirai, S.~Kumano and T.~-H.~Nagai,
  Phys.\ Rev.\ C {\bf 76}, 065207 (2007).

\bibitem{deFlorian:2011fp} 
  D.~de Florian, R.~Sassot, P.~Zurita and M.~Stratmann,
  Phys.\ Rev.\ D {\bf 85}, 074028 (2012).

\bibitem{Kovarik:2010uv} 
  K.~Kovarik, I.~Schienbein, F.~I.~Olness {\it et al.}, 
  Phys.\ Rev.\ Lett.\  {\bf 106}, 122301 (2011).
  
\bibitem{Paukkunen:2013grz} 
  H.~Paukkunen and C.~A.~Salgado,
  Phys.\ Rev.\ Lett.\  {\bf 110}, 212301 (2013).


\bibitem{KP04}
  S.~A.~Kulagin and R.~Petti,
  Nucl.\ Phys.\  A {\bf 765}, 126 (2006).

\bibitem{KP07}
  S.~A.~Kulagin and R.~Petti,
  Phys.\ Rev.\  D {\bf 76}, 094023 (2007).

\bibitem{KP10} 
  S.~A.~Kulagin and R.~Petti,
  Phys.\ Rev.\ C {\bf 82}, 054614 (2010).

\bibitem{Akulinichev:1985ij} 
  S.~V.~Akulinichev, S.~A.~Kulagin and G.~M.~Vagradov,
  Phys.\ Lett.\ B {\bf 158}, 485 (1985).

\bibitem{Akulinichev:1986gt} 
   S.~V.~Akulinichev, G.~M.~Vagradov and S.~A.~Kulagin,
   JETP Lett.\  {\bf 42}, 127 (1985)
   [Pisma Zh.\ Eksp.\ Teor.\ Fiz.\  {\bf 42}, 105 (1985)].

\bibitem{Akulinichev:1985xq} 
   S.~V.~Akulinichev, S.~Shlomo, S.~A.~Kulagin and G.~M.~Vagradov,
   Phys.\ Rev.\ Lett.\  {\bf 55}, 2239 (1985).

\bibitem{Ku89}
  S.~A.~Kulagin,
  Nucl.\ Phys.\  A {\bf 500}, 653 (1989).

\bibitem{Kulagin:1994fz} 
  S.~A.~Kulagin, G.~Piller and W.~Weise,
  Phys.\ Rev.\ C {\bf 50}, 1154 (1994).

\bibitem{Ku98}
  S.~A.~Kulagin,
  Nucl.\ Phys.\  A {\bf 640}, 435 (1998).

\bibitem{Kulagin:2008fm} 
  S.~A.~Kulagin and W.~Melnitchouk,
  Phys.\ Rev.\ C {\bf 78}, 065203 (2008).
  

\bibitem{Bauer:iq}
T.~H.~Bauer, R.~D.~Spital, D.~R.~Yennie and F.~M.~Pipkin,
Rev.\ Mod.\ Phys.\  {\bf 50}, 261 (1978)
[Erratum-ibid.\  {\bf 51}, 407 (1979)].

\bibitem{Nikolaev:1990ja} 
  N.~N.~Nikolaev and B.~G.~Zakharov,
  Z.\ Phys.\ C {\bf 49}, 607 (1991).

\bibitem{Piller:1999wx} 
  G.~Piller and W.~Weise,
  Phys.\ Rept.\  {\bf 330}, 1 (2000).

\bibitem{Glauber:1955qq}
R.~J.~Glauber,
Phys.\ Rev.\  {\bf 100}, 242 (1955).

\bibitem{Gribov:1968gs}
V.~N.~Gribov,
Sov.\ Phys.\ JETP {\bf 30}, 709 (1970)
[Zh.\ Eksp.\ Teor.\ Fiz.\  {\bf 57}, 1306 (1969)].


\bibitem{Nikolaev:1975vy} 
  N.~N.~Nikolaev and V.~I.~Zakharov,
  Phys.\ Lett.\ B {\bf 55}, 397 (1975).

\bibitem{Frankfurt:1990xz} 
  L.~L.~Frankfurt, M.~I.~Strikman and S.~Liuti,
  Phys.\ Rev.\ Lett.\  {\bf 65}, 1725 (1990).

\bibitem{Ioffe:1985ep} 
  B.~L.~Ioffe, V.~A.~Khoze and L.~N.~Lipatov,
  ``Hard Processes. Vol. 1: Phenomenology, Quark Parton Model,''
  Amsterdam, Netherlands: North-Holland (1984).

\bibitem{Kulagin:1998wc}
S.~A.~Kulagin,
arXiv:hep-ph/9812532.

\bibitem{LlewellynSmith:1983qa}
C.~H.~Llewellyn Smith,
Phys.\ Lett.\ B {\bf 128}, 107 (1983).

\bibitem{Ericson:1983um}
M.~Ericson and A.~W.~Thomas,
Phys.\ Lett.\ B {\bf 128}, 112 (1983).

\bibitem{Berger:1984na}
E.~L.~Berger and F.~Coester,
Phys.\ Rev.\ D {\bf 32}, 1071 (1985).

\bibitem{Friman:1983rt}
B.~L.~Friman, V.~R.~Pandharipande and R.~B.~Wiringa,
Phys.\ Rev.\ Lett.\  {\bf 51}, 763 (1983).

\bibitem{Sapershtein:1985pa}
E.~E.~Sapershtein and M.~Z.~Shmatikov,
JETP Lett.\  {\bf 41}, 53 (1985)
[Pisma Zh.\ Eksp.\ Teor.\ Fiz.\  {\bf 41}, 44 (1985)].

\bibitem{Kaptari:1989un}
L.~P.~Kaptari {\it et. al.},
Nucl.\ Phys.\ A {\bf 512}, 684 (1990).

\bibitem{Jung:1990pu}
H.~Jung and G.~A.~Miller,
Phys.\ Rev.\ C {\bf 41}, 659 (1990).

\bibitem{Koltun:1997py}
D.~S.~Koltun,
Phys.\ Rev.\ C {\bf 57}, 1210 (1998).

\bibitem{Korpa:2013ia} 
  C.~L.~Korpa and A.~E.~L.~Dieperink,
  Phys.\ Rev.\ C {\bf 87}, 014616 (2013).

  
\bibitem{Alekhin:2006zm} 
  S.~Alekhin, K.~Melnikov and F.~Petriello,
  Phys.\ Rev.\ D {\bf 74}, 054033 (2006).

\bibitem{Alekhin:2007fh} 
  S.~Alekhin, S.~A.~Kulagin and R.~Petti,
  AIP Conf.\ Proc.\  {\bf 967}, 215 (2007).
  

\bibitem{Bickerstaff:1985ax}
  R.~P.~Bickerstaff, M.~C.~Birse and G.~A.~Miller,
  Phys.\ Rev.\ Lett.\  {\bf 53}, 2532 (1984).


\bibitem{McGaughey:1999mq} 
  P.~L.~McGaughey, J.~M.~Moss and J.~C.~Peng,
  Ann.\ Rev.\ Nucl.\ Part.\ Sci.\  {\bf 49}, 217 (1999).
  
\bibitem{Peng:2014hta} 
  J.~-C.~Peng and J.~-W.~Qiu,
  Prog.\ Part.\ Nucl.\ Phys.\  {\bf 76}, 43 (2014).

\bibitem{E772}
  D.~M.~Alde {\it et al.},
  Phys.\ Rev.\ Lett.\  {\bf 64}, 2479 (1990).

\bibitem{E866} 
  M.~A.~Vasilev {\it et al.},  
  Phys.\ Rev.\ Lett.\  {\bf 83}, 2304 (1999).
  
\bibitem{dy-web}
E866/E789/E772 web resources 
\url{http://p25ext.lanl.gov/e866/papers/papers.html}

\bibitem{Londergan:2009kj} 
  J.~T.~Londergan, J.~C.~Peng and A.~W.~Thomas,
  Rev.\ Mod.\ Phys.\  {\bf 82}, 2009 (2010).
    

\bibitem{Garvey:2002sn} 
  G.~T.~Garvey and J.~C.~Peng,
  Phys.\ Rev.\ Lett.\  {\bf 90}, 092302 (2003).

\bibitem{Accardi:2009qv} 
  A.~Accardi, F.~Arleo, W.~K.~Brooks {\it et al.}, 
  Riv.\ Nuovo Cim.\  {\bf 32}, 439 (2010).


\bibitem{E906}
  J. Arrington, {\it et al.}, [E906 Collaboration], 
  Fermilab Projects-doc-395, 2006,
  \url{http://projects-docdb.fnal.gov/cgi-bin/ShowDocument?docid=395};
  P.~E.~Reimer, 
  Eur.\ Phys.\ J.\ A {\bf 31}, 593 (2007); 
  P.~E.~Reimer [Fermilab SeaQuest Collaboration],
  J.\ Phys.\ Conf.\ Ser.\ {\bf 295}, 012011 (2011).  

\bibitem{Kulagin:2003wz}
S.~A.~Kulagin,
Phys.\ Rev.\ D {\bf 67}, 091301 (2003).

\bibitem{Kulagin:2004jn}
S.~A.~Kulagin,
Nucl.\ Phys.\ Proc.\ Suppl.\  {\bf 139}, 213 (2005).

\bibitem{Mishra:2008nx} 
  S.~R.~Mishra, R.~Petti and C.~Rosenfeld,
  PoS NUFACT {\bf 08}, 069 (2008).

\bibitem{Adams:2013qkq} 
  C.~Adams {\it et al.}  (LBNE Collaboration),
  arXiv:1307.7335 [hep-ex].

\bibitem{Cloet:2009qs} 
  I.~C.~Cloet, W.~Bentz and A.~W.~Thomas,
  Phys.\ Rev.\ Lett.\  {\bf 102}, 252301 (2009).

\bibitem{Subedi:2008zz} 
  R.~Subedi {\it et al.}, 
  Science {\bf 320}, 1476 (2008).

\bibitem{Sargsian:2012sm} 
  M.~M.~Sargsian,
  Phys.\ Rev.\ C {\bf 89}, 034305 (2014).

\bibitem{Moreno:1990sf} 
  G.~Moreno, C.~N.~Brown, W.~E.~Cooper {\it et al.},
  Phys.\ Rev.\ D {\bf 43}, 2815 (1991).


\end{thebibliography}
\end{document}